\documentclass[11pt]{article}
\usepackage{amsmath,amsfonts,latexsym, amssymb}

\usepackage{color}

\newcommand{\wt}{\widetilde}
\newcommand{\wh}{\widehat}

\newcommand{\beqa}{\begin{eqnarray}}
\newcommand{\eeqa}{\end{eqnarray}}
\newcommand{\e}{\varepsilon}
\newcommand{\eps}{\varepsilon}
\newcommand{\pt}{\partial}
\newcommand{\rd}{{\rm d}}
\newcommand{\bR}{{\mathbb R}}

\newcommand{\non}{\nonumber}

\newcommand{\tr}{\mbox{Tr\,}}

\newcommand{\bm}{{\bf{m}}}
\newcommand{\ba}{{\bf{a}}}

\newcommand{\bx}{{\bf{x}}}
\newcommand{\by}{{\bf{y}}}
\newcommand{\bu}{{\bf{u}}}
\newcommand{\bv}{{\bf{v}}}

\newcommand{\mg}{{m_N}}
\newcommand{\mW}{{m_W}}

\newcommand{\al}{\alpha}

\newcommand{\be}{\begin{equation}}
\newcommand{\ee}{\end{equation}}

\newcommand{\om}{{\omega}}

\newcommand{\cE}{{\mathcal E}}
\newcommand{\cG}{{\mathcal G}}

\newcommand{\cB}{{\mathcal B}}

\newcommand{\cN}{{\mathcal N}}
\newcommand{\cH}{{\mathcal H}}

\newcommand{\E}{{\mathbb E }}
\newcommand{\R}{{\mathbb R }}
\newcommand{\N}{{\mathbb N}}

\renewcommand{\P}{{\mathbb P}}
\newcommand{\C}{{\mathbb C}}
\newcommand{\bH}{{\mathbb H}}

\newtheorem{theorem}{Theorem}
\newtheorem{corollary}[theorem]{Corollary}
\newtheorem{lemma}[theorem]{Lemma}
\newtheorem{proposition}{Proposition}

\newcommand{\qed}{\hfill\fbox{}\par\vspace{0.3mm}}

\numberwithin{equation}{section}
\numberwithin{theorem}{section}
\numberwithin{definition}{section}
\numberwithin{proposition}{section}
\numberwithin{remark}{section}

\title{The local relaxation flow approach to universality of the local
statistics for random matrices}

\author{
L\'aszl\'o Erd\H os${}^1$\thanks{Partially supported
by SFB-TR 12 Grant of the German Research Council},
Benjamin Schlein${}^2$\;
Horng-Tzer Yau${}^3$\thanks{Partially supported
by NSF grants DMS-0757425, 0804279}  \; and Jun Yin${}^3$ \\ \\
Institute of Mathematics, University of Munich, \\
Theresienstr. 39, D-80333 Munich, Germany \\ lerdos@math.lmu.de ${}^1$ \\ \\
Department of Pure Mathematics and Mathematical Statistics
\\  University of Cambridge \\
Wilberforce Rd, Cambridge CB3 0WB, UK\\ b.schlein@dpmms.cam.ac.uk ${}^2$ \\ \\
Department of Mathematics, Harvard University\\
Cambridge MA 02138, USA \\  htyau@math.harvard.edu,  jyin@math.harvard.edu ${}^3$ \\ \\
\\}

\begin{document}

\date{Aug 17, 2010}

\maketitle

\begin{abstract}

We present a generalization of the method of the local relaxation
flow to establish the universality of local spectral statistics
of a broad class of large random matrices. We show that
the local  distribution of the eigenvalues
coincides with the local statistics of the corresponding  Gaussian ensemble
provided the distribution of the individual matrix element is smooth and
the eigenvalues  $\{ x_j\}_{j=1}^N$ are close to their
classical location $\{ \gamma_j\}_{j=1}^N$ determined by the limiting density of
eigenvalues. Under the scaling where the typical distance between neighboring
eigenvalues is of order $1/N$, the necessary apriori estimate on the
location of eigenvalues requires only to know that $\E |x_j - \gamma_j |^2 \le
N^{-1-\e}$ on average. This information can be obtained by well established
methods for various matrix ensembles. We demonstrate the method by
proving local spectral universality for sample covariance matrices.

\bigskip

Nous pr\'esentons une g\'en\'eralisation de la m\'ethode du {\it flot
de relaxation locale} servant \`a \'etablir l'universalit\'e des
statistiques spectrales locales d'une vaste classe de grandes matrices
al\'eatoires. Nous d\'emontrons que la distribution locale des valeurs
propres co\"incide avec celle de l'ensemble gaussien
pourvu que la loi des coefficients individuels de la matrice soit lisse
et que les valeurs propres $\{ x_j\}_{j=1}^N$ soient pr\`es de leurs
quantiles classiques $\{ \gamma_j\}_{j=1}^N$ determin\'ees par la densit\'e
limite des valeurs
propres. Dans la normalisation o\`u la distance typique
entre les valeurs propres voisines est d'ordre $1/N$, la borne
a priori n\'ecessaire sur la position des valeurs propres n\'ecessite
uniquement l'\'etablissement de $\E |x_j - \gamma_j |^2 \le N^{-1-\e}$
en moyenne.
Cette information peut \^etre obtenue par des m\'ethodes bien \'etablies
pour divers ensembles de matrices.
Nous illustrons la m\'ethode en d\'emontrant l'universalit\'e
spectrale locale pour des matrices de covariance.

\end{abstract}


\medskip

{\it Running title:} Local relaxation flow

\medskip


\medskip


\section{Introduction}

A central question concerning random matrices is the universality conjecture
which states that  local statistics  of eigenvalues of large
$N\times N$ square matrices $H$
are determined by  the symmetry type of the ensembles
but are otherwise  independent of the details of  the distributions.
In particular they coincide with that of the corresponding
Gaussian ensemble. The most commonly studied ensembles
are

(i) hermitian, symmetric and quaternion self-dual matrices with identically
distributed and centered entries
that are independent (subject to the natural restriction of the symmetry);

(ii) sample covariance matrices of the form $H=A^*A$, where
$A$ is an $M \times N$ matrix with centered real or complex i.i.d. 
entries.

There are two types of universalities: the edge universality and the
bulk universality concerning energy levels near the spectral
edges and in the interior of the spectrum, respectively.  Since the
works of Sinai and Soshnikov \cite{SS, Sosh}, the edge
universality is commonly approached via the fairly robust
moment method \cite{P, FSo, So1, Sosh2, Ruz}; very recently an alternative
approach was given in \cite{TV2}.

The bulk universality is a subtler
problem. In the simplest case of the hermitian Wigner ensemble,
it states that, independent of the distribution of the entries,
the local $k$-point correlation functions
of the eigenvalues (see \eqref{corr} for the precise definition later), after appropriate
rescaling and in the $N\to\infty$ limit, are given
by the determinant of the  {\it sine kernel}
\be
\det \big( K(x_\ell - x_j)\big)_{\ell,j=1}^k, \qquad K(x) = \frac{\sin \pi x}{\pi x}.
\label{sine}
\ee
Similar statement is expected to hold for all other ensembles mentioned above
but the explicit formulas are somewhat more complicated. Detailed formulas
for the different Wigner ensembles can be found
e.g., in \cite{M}. The various sample covariance ensembles have the same local statistics
for their {\it singular values} as the local eigenvalue
statistics of the corresponding Wigner ensembles.

For ensembles of hermitian, symmetric or quaternion self-dual matrices
that remain invariant under the
transformations $H\to U^*HU$ for any unitary, orthogonal or
symplectic matrix $U$, respectively, the joint
probability density function of all the $N$ eigenvalues can be explicitly
computed.  These ensembles are typically given by the
probability density
\be
  P(H)\rd H \sim \exp(-N\tr V(H)) \rd H,
\label{inv}
\ee
where $V$ is a real function with sufficient growth at
infinity and $\rd H$ is the flat Lebesgue measure on the
corresponding symmetry class of matrices.
The eigenvalues are strongly correlated and they
are distributed according to a Gibbs measure
with a long range logarithmic interaction potential.
The joint probability density of the eigenvalues of $H$
with distribution \eqref{inv}
can be computed explicitly:
\be
  f(x_1, x_2, \ldots, x_N)
= \mbox{(const.)} \prod_{i<j} |x_j-x_i|^\beta \prod_{j=1}^N
 e^{- N\sum_{j=1}^N V(x_j)},
\label{expli}
\ee
where $\beta=1,2, 4$ for hermitian, symmetric and  symplectic ensembles,
respectively, and $\mbox{const.}$ is a normalization factor.
The formula \eqref{expli} defines a joint probability
density of $N$ real random variables for any $\beta\ge 1$ even
when there is no underlying matrix ensemble. This ensemble
is called the {\it invariant $\beta$-ensemble}.
Quadratic $V$ corresponds to the Gaussian  ensembles; we note that these are
the only ensembles that are simultaneously invariant and have i.i.d. matrix entries. These are called the Gaussian Orthogonal, Unitary
and Symplectic Ensembles (GOE, GUE, GSE for short) in case of $\beta=1,2,4$, respectively. Somewhat different choices of $V$ lead
to two other classical ensembles, the Laguerre and the Jacobi ensembles,
that also have matrix interpretation for $\beta=1,2, 4$ (e.g., the Laguerre ensemble
corresponds to the Gaussian sample covariance matrices which are
also called {\it Wishart matrices}), see \cite{DE, For} for more details.
The local statistics can be obtained via a detailed analysis of
orthogonal polynomials on the real line with respect to the
weight function $\exp(-V(x))$.  This approach was originally applied
to classical ensembles by Dyson \cite{D}, Mehta and Gaudin \cite{MG} and Mehta \cite{M}
that lead to  classical orthogonal polynomials.
Later general methods using orthogonal polynomials were developed to tackle a very general
class of invariant ensembles by Deift {\it et.al.}, see
\cite{De1, De2, DKMVZ1, DKMVZ2} and references therein,
and also by Bleher and Its \cite{BI} and Pastur and Schcherbina \cite{PS}.

Many natural matrix ensembles are typically not unitarily invariant; the most
prominent examples are the Wigner matrices or
the sample covariance matrices mentioned in  (i) and (ii).
For these ensembles, apart from the identically distributed Gaussian case,
no explicit formula is available for the joint eigenvalue
distribution. Thus the basic algebraic connection between
eigenvalue ensembles and orthogonal polynomials is missing
and completely new methods needed to be developed.

The bulk universality for  {\it hermitian} Wigner ensembles
has been established recently in \cite{ERSY2}, by Tao and Vu in
\cite{TV} and  in \cite{ERSTVY}.
These works rely on the Wigner
matrices with Gaussian divisible distribution, i.e.,
ensembles  of the form
\be
\wh H+ \sqrt{s} V,
\label{HaV}
\ee
where $\wh H$  is a Wigner matrix,
$V$ is an independent standard GUE matrix and $s$ is a positive constant.
Johansson  \cite{J} (see also Ben Arous and P\'ech\'e \cite{BP}
and the recent paper  \cite{J1}) proved the bulk universality for
the eigenvalues of such matrices
by an asymptotic analysis on an {\it explicit} formula for
the correlation functions adapted from
Br\'ezin-Hikami  \cite{BH}.
Unfortunately, the similar formula for symmetric or quaternion self-dual Wigner matrices,
as well as for real sample covariance matrices,
is not very explicit and the technique of \cite{BP, ERSY2, J}
cannot be extended to prove universality.
Complex sample covariance matrices
can however be handled with an analogous formula \cite{BP} and universality
without any Gaussian component is a work in progress \cite{BP1}.

A key observation of Dyson is that if the matrix $\wh H+ \sqrt{s} V$
is embedded into a stochastic matrix flow, i.e. one considers
$\wh H + V(s)$ where the matrix elements of $V(s)$ are 
independent standard Brownian motions with variance $s/N$, then
the evolution of the eigenvalues is given by a  system of coupled
stochastic differential equations (SDE), commonly called
the  Dyson Brownian motion (DBM) \cite{Dy}.
If we replace the Brownian motions  by the
Ornstein-Uhlenbeck  processes to keep the variance constant,
then the resulting dynamics on the eigenvalues,
which we still call
DBM,  has the GUE eigenvalue  distribution as the invariant measure.
Similar stochastic processes can be constructed for symmetric, quaternion self-dual
and sample covariance type matrices, and, in fact, on the level of
eigenvalue SDE they can be
extended to other values of $\beta$ (see \eqref{sde} and \eqref{sdeW2} for
the precise formulas).

The result of \cite{J, BP} can be interpreted as stating that the
local statistics of GUE is
reached via DBM for time of order one. In fact, by analyzing the dynamics
of DBM with ideas from the hydrodynamical limit,
we have extended Johansson's result  to $s\gg N^{-3/4}$  \cite{ERSY}.
The  key
observation of \cite{ERSY}   is that the local statistics of
eigenvalues  depend exclusively on the approach to local equilibrium
which in general is faster than reaching the global equilibrium.
Unfortunately, the identification of  local equilibria  in \cite{ERSY} still uses
explicit representations of correlation functions by  orthogonal polynomials
(following e.g. \cite{PS}),   and the extension to other ensembles is not a simple task.

In \cite{ESY4} we introduced an approach 
based on a new stochastic flow,  the {\it local relaxation flow}, which locally behaves like
DBM, but has a faster decay to equilibrium.  This method 
completely circumvented explicit formulas and  it resulted in
proving universality for {\it symmetric} Wigner matrices
(the method applies to hermitian and quaternion self-dual Wigner matrices as well).
As an input of this method,
we needed a fairly detailed control on the local density of
eigenvalues that could be obtained from our previous works on Wigner
matrices \cite{ESY1, ESY2, ESY3}.

In this paper we will prove a general theorem which states
that as long as the eigenvalues are  at most $N^{-1/2-\e}$ distance
near their classical location on average,
the local statistics is universal and in particular it coincides
with the Gaussian case for which explicit formulas have been
computed.   To introduce  this flow, denote by  $\gamma_j$  the location of the $j$-th eigenvalue  that will be
defined in \eqref{gammaj}. We first define the 
 {\it pseudo equilibrium  measure}  by 
\be
\om_N=C_N \exp \big(-NW \big) \mu_N,  \quad W(\bx) =    \sum_{j=1}^N
W_j (x_j)   , \qquad W_j (x) = \frac{1}{2R^2} (x_j -\gamma_j)^2,
\label{pseudo}
\ee
where $\mu_N$ is the probability measure for the eigenvalue distribution 
of the corresponding  Gaussian ensemble. In case of Wigner matrices, $\mu_N$  is the measure 
for the general $\beta$ ensemble ($\beta\ge 1$  and $\beta=2$ for GUE):
\be\label{Huj}
\mu=\mu_N(\rd{\bf x})=
\frac{e^{-\cH({\bf x})}}{Z_\beta}\rd{\bf x},\qquad \cH({\bf x}) =
N \left [ \beta \sum_{i=1}^N \frac{x_{i}^{2}}{4} -  \frac{\beta}{N} \sum_{i< j}
\log |x_{j} - x_{i}| \right ].
\ee 
In this setting, it is natural to view eigenvalues 
as  random points and their  equilibrium measure as Gibbs measure
with a Hamiltonian $\cH$. We will freely use the
 terminology of statistical mechanics.  
Note that the additional term $W_j$ in $\om_N$ confines
the $j$-th point $x_j$ near its classical location, but  the probability w.r.t. the 
equilibrium measure $\mu_N$ of the event that 
$x_j$ near its classical location will be shown to be  very  close to $1$. Furthermore, 
we will prove that the local statistics of the measures $\om_N$ and $\mu_N$ 
are identical in the limit $N \to \infty$ and this justifies  the term pseudo equilibrium  measure.

 The local relaxation flow is defined to be the reversible flow
 (or the gradient flow) generated by the 
pseudo-equilibrium measure. The main advantage of the local relaxation flow is that it   has a faster
 decay to global  equilibrium   (Theorem \ref{thm2})  compared with the DBM.  The idea behind
 this construction can be related to the treatment of metastability in statistical physics.
 Imagine that we have a double well potential and we wish to treat the 
dynamics of a particle in one of the two wells. Up to a certain time, say $t_0$,  the particle 
 will be confined in the  well where the particle initially located. However, the potential
 of this particle, given by the double well, is not convex. A naive idea is to regain the 
convexity before  the time $t_0$  is to modify the potential to be a single well! Now as
 long as we can prove that the particle was confined in the initial well up to $t_0$, there 
is no difference between these two dynamics. But the modified dynamics, being w.r.t. a 
convex potential, can be estimated much more precisely and this estimate can be carried
 over to the original dynamics up to the time $t_0$.

In our case, the convexity of the equilibrium measure $\mu_N$ is rather weak and in fact, it comes 
from the quadratic confining potential $\beta x_i^2/4$ of \eqref{Huj}.
 So the potential is convex, just not ``convex enough".
There is no sharp transition like jumping from one metastable state to another
as in the double well case.  Instead,  there are  two time
scales: in short time the local equilibrium is formed, on longer time, it approaches
 the global equilibrium. 
The approach to the local equilibrium is governed by 
a strong  intrinsic convexity in certain directions 
due to  the interactions 
(see \eqref{convex} later for a precise formula). 
To reveal this additional convexity, in our previous paper \cite{ESY4}
we introduced a pseudo equilibrium measure where we
replaced the long range part of the interaction by a  mean-field
potential term using the classical locations of far away particles.
This potential term inherited the intrinsic convexity of the
interaction and it could be directly used to enhance the decay 
to the local statistics. 
One technical difficulty with this approach was that we needed to handle
the singular behavior of the logarithmic interaction potential.
In this paper we show that the pseudo equilibrium measure
can be defined by adding a Gaussian term. This simple
modification turns out to be sufficient and  is also model-independent.
Since the Gaussian modification is regular, we no longer need 
to deal with singularities. The price to pay is that we need a 
slightly stronger local semicircle law which will be treated in 
Section \ref{sec:wish}.

The method of local relaxation flow itself proves universality for
Wigner matrices with a small Gaussian component $\sqrt{s}V$ (typically of
variance $s\ge N^{-\gamma}$ with some $0<\gamma <1$). 
In other words, we can prove  universality for a Wigner ensemble
whose single entry distribution (the distribution of its matrix elements)
is given by $e^{t B} u_0$, where  $B$ is the generator of the 
Ornstein-Uhlenbeck process and $u_0$ is any initial distribution (We remark that 
in our approach of decay to equilibrium, the Brownian motion in the
construction of DBM is always replaced by  
the Ornstein-Uhlenbeck process). 
To obtain
universality for Wigner matrices without any Gaussian component, it remains to prove that 
for a given Wigner matrix ensemble with a single entry distribution $\nu$ we can find 
$u_0$ and $t$ such that the eigenvalue distributions  of the ensembles given by 
$\nu$ and $e^{t B} u_0$ are very close to each other. By  the {\it method of
reverse heat flow} introduced in \cite{ERSY2},  we choose  $u_0$ to be  
an approximation of  $e^{-t B} \nu$. Although the  Ornstein-Uhlenbeck evolution cannot be reversed, 
we can approximately reverse it provided that  $\nu$ is sufficient smooth
and the  time is short. 
This enables us to compare local statistics
of Wigner ensembles with and without small Gaussian components
assuming that the single entry distribution is sufficiently smooth
(see Section \ref{sec:rev}).

As an application, we will use  this method to prove  the bulk universality of  
sample covariance ensembles. 
The necessary apriori control on the location
of eigenvalues will be obtained by a local semicircle law.  In addition to 
 sample covariance ensembles,  we will outline  
the modifications needed for proving the bulk 
universality of  symplectic ensembles.

\section{Universality for the local relaxation flow}\label{sec:flow}

In this section, we consider the following general setup.
Suppose $\mu = e^{-N\cH} /Z $ is a probability measure on the configuration
space $\bR^N$ characterized by some Hamiltonian $\cH :\bR^N\to \bR$, where
$Z= \int e^{-N\cH(\bx)} \rd \bx<\infty$ is the normalization.
We will always assume that $\cH$ is symmetric under permutation of the variables
$\bx =(x_1, x_2, \ldots, x_N)\in \bR^N$.

We  consider time dependent permutational symmetric
probability measures with density $f_t(\bx)$, $t\ge0$,
with respect to the measure $\mu(\rd \bx)=\mu(\bx)\rd\bx$.
The dynamics is characterized by the forward equation
\be\label{dy}
\partial_{t} f_t =  L f_t, \qquad t\ge 0,
\ee
with a given permutation symmetric initial data $f_0$.
Here the generator $L$ is defined via
the Dirichlet form as
\be
D(f) = D_\mu(f) =  -\int  f L f  \rd \mu =  \sum_{j=1}^N \frac{1}{2N}
\int (\partial_j f)^2 \rd \mu,  \quad \pt_j =\pt_{x_j}.
\label{def:dir2}
\ee
Formally, we have $L= \frac{1}{2N}\Delta - \frac{1}{2}(\nabla \cH)\nabla$.
In Appendix \ref{sec:defdyn} we will
show that under general conditions on $\cH$ the generator can be defined as a self-adjoint
operator on an appropriate domain and the dynamics is well defined for any $f_0\in L^1(\rd\mu)$
initial data.
Strictly speaking, we will consider a sequence of Hamiltonians $\cH_N$ and corresponding
dynamics $L_N$ and $f_{t,N}$
parametrized by $N$, but the $N$-dependence will be omitted. All results
will concern the $N\to\infty$ limit.

\bigskip

The expectation with respect to the  density $f_t$ will be denoted by $\E_t$ with
$\E:= \E_0$. The expectation with respect to the equilibrium measure  $\mu$ is denoted by $\E^\mu$.
For any $n\ge 1$ we define the $n$-point
correlation functions (marginals) of the probability measure $f_t\rd\mu$ by
\be
 p^{(n)}_{t,N}(x_1, x_2, \ldots,  x_n) = \int_{\R^{N-n}}
f_t(\bx) \mu(\bx) \rd x_{n+1}\ldots
\rd x_N.
\label{corr}
\ee
With a slight abuse of notations, we will sometimes  also use $\mu$ to denote
the density of the measure $\mu$ with respect to the Lebesgue measure.
The correlation functions of the equilibrium measure  are denoted by
$$
 p^{(n)}_{\mu,N}(x_1, x_2, \ldots,  x_n) = \int_{\R^{N-n}}
\mu(\bx) \rd x_{n+1}\ldots
\rd x_N.
$$

We now list our main assumptions on the initial distribution $f_0$
and on its evolution $f_t$.
We first define the subdomain
\be
\Sigma_N: = \big\{ \bx\in\bR^N, \; x_1 < x_2 < \ldots < x_N\big\}
\label{def:Sigma}
\ee
of ordered sets of points $\bx$. In the application to the
sample covariance matrices, we will use the subdomain
\be
\Sigma_N^+: = \big\{ \bx\in\bR^N, \; 0<x_1 < x_2 < \ldots < x_N\big\}
\label{def:SigmaW}
\ee
of ordered sets of positive points.

\bigskip

{\bf Assumption I.} The Hamiltonian $\cH$ of the equilibrium measure has the
form
\be
  \cH = \cH_N(\bx)  = \beta\Big[ \sum_{j=1}^N U(x_j) 
-\frac{1}{N}\sum_{i<j} \log|x_i-x_j| \Big], 
\label{ham}
\ee
where $\beta\ge 1$. The function $U:\bR\to \bR$ is smooth with
$U'' \ge 0$  and 
\be\label{U}
  U(x) \ge C  |x|^\delta  \qquad  \text{ for some $\delta> 0$ and $|x|$ large} .
\ee
The condition $U''\ge 0$ can be relaxed to $\inf U''>-\infty$, see 
remark after  \eqref{UprimeM}.

\medskip

Alternatively, in order to discuss the case of  the
sample covariance matrices, we will
also consider the following modification of Assumption I.

\bigskip

{\bf Assumption I'.} The Hamiltonian $\cH$ of the equilibrium measure has the
form
\be
  \cH = \cH_N(\bx)  = \beta\Big[ \sum_{j=1}^N U(x_j) -
\frac{1}{N}\sum_{i<j} \log|x_i-x_j|  - \frac{1}{N}\sum_{i<j} \log|x_i+x_j|-
\frac{c_N}{N}\sum_j \log |x_j|\Big], 
\label{hamW}
\ee
where $\beta\ge 1$ and  $c_N\ge 1$. The function $U$ satisfies the same
conditions as in Assumption I.

\medskip
It is easy to check that the condition \eqref{U} guarantees that the following
bound holds for the normalization constant
\be
    |\log Z|\le CN^m
\label{Zbound}
\ee
with some exponent $m$ depending on $\delta$.
\bigskip

In Appendix \ref{sec:defdyn} we will show that for $\beta\ge 1$ the dynamics
\eqref{dy} can be restricted to the subdomains $\Sigma_N$ or $\Sigma_N^+$,
respectively, i.e,
the ordering will be preserved under the dynamics.
In the sequel we will thus assume
that $f_t$ is a probability measure on $\Sigma_N$ or $\Sigma_N^+$.
We continue to use the notation $f$
and $\mu$ for the restricted measure. Note that the correlation
functions $p^{(k)}$ from \eqref{corr} are still defined on $\bR^k$, i.e., their
arguments remain unordered. 

\medskip

It follows from Assumption I (or I') that the Hessian matrix of $\cH$ satisfies the
following bound: 
\be
 \big\langle \bv, \nabla^2 \cH(\bx) \bv\big\rangle
\ge  \frac{\beta}{N}
\sum_{i<j} \frac{ (v_i-v_j)^2}{(x_i-x_j)^2}, \qquad
\bv = (v_1, \ldots , v_N)\in\bR^N, \quad \bx\in \Sigma_N \quad (\mbox{or}
\;\bx \in \Sigma_N^+).
\label{convex}
\ee

This convexity bound is the key assumption;
our method works for a broad class of  general  Hamiltonians
as long as \eqref{convex} holds.
In particular, an arbitrary many-body potential function $V(\bx)$
can be added to the Hamiltonians \eqref{ham}, \eqref{hamW},
as long as $V$ is convex on the open sets $\Sigma_N$ and $\Sigma_N^+$,
respectively. The argument in the proof of the main Theorem
\ref{thm:main} remains unchanged, but the technical
details of the regularization of the singular dynamics
(Appendix \ref{sec:BE}) becomes more involved. We do not
pursue this direction here since we do not need it
for the application for Wigner and sample covariance matrices.

\bigskip

{\bf Assumption II.} There exists a continuous, compactly supported
density function $\varrho(x)\ge 0$, $\int_\bR \varrho =1$, on the
real line, independent of $N$, such that for any fixed $a,b\in \bR$
\be
\lim_{N\to\infty}   \sup_{t\ge 0}  \Bigg|
\int \frac{1}{N}\sum_{j=1}^N {\bf 1} ( x_j \in [a, b]) f_t(\bx)\rd\mu(\bx)
- \int_a^b \varrho(x) \rd x \Bigg| =0.
\label{assum1}
\ee

\bigskip

Let $\gamma_j =\gamma_{j,N}$ denote the location of the $j$-th point
under the limiting density, i.e., $\gamma_j$ is defined by
\be
 N \int_{-\infty}^{\gamma_j} \varrho(x) \rd x = j, \qquad 1\leq j\le N, \quad
\gamma_j\in \mbox{supp} \varrho.
\label{gammaj}
\ee
We will call $\gamma_j$ the {\it classical location} of the $j$-th point.
Note that $\gamma_j$ may not be uniquely defined if the support
of $\varrho$ is not connected
but in this case the next Assumption III
will not be satisfied anyway.

\bigskip

{\bf Assumption III.} There exists an $\e>0$ such that
\be
 \sup_{t\ge N^{-2\e}} \int  \frac{1}{N}\sum_{j=1}^N(x_j-\gamma_j)^2
 f_t(\rd \bx)\mu(\rd \bx) \le CN^{-1-2\e}
\label{assum2}
\ee
with a constant $C$ uniformly in $N$.

\bigskip

Under
Assumption II, the typical spacing between
neighboring points is of order $1/N$ away from the spectral edges, i.e., in
the vicinity of any energy $E$ with $\varrho(E)>0$. Assumption III guarantees
that typically the random points $x_j$ remain in
the $N^{-1/2-\e}$ vicinity of their classical location.

\bigskip

The final assumption is an upper bound on the local density. 
For any $I\in \R$, let
$$
    \cN_I: = \sum_{i=1}^N {\bf 1}( x_i \in I)
$$
denote the number of points in $I$.

\bigskip

{\bf Assumption IV.} For any compact subinterval $I_0\subset \{ E\;: \; \varrho(E)>0\}$,
and for  any $\delta>0$,  $\sigma>0$
there are constants $C_n$, $n\in \N$, depending on $I_0$,
 and $\sigma$ such that for any interval $I\subset I_0$ with
$|I|\ge N^{-1+\sigma}$ and for any $K\ge 1$, we have
\be
   \sup_{\tau \ge N^{-2\e}}
    \int {\bf 1}\big\{ \cN_I \ge KN|I| \big\}f_\tau \rd\mu
\le C_{n} K^{-n}, \qquad n=1,2,\ldots,
\label{ass4}
\ee
where $\e$ is the exponent from Assumption III.
\bigskip

The main general theorem is the following:

\begin{theorem}\label{thm:main}  Suppose that the Hamiltonian given in
\eqref{ham} or \eqref{hamW} satisfy Assumption I or I', respectively.
Suppose that
Assumptions II, III  and IV hold for the solution $f_t$ of the forward equation
\eqref{dy}.  Assume that at time $t_0=N^{-2\e}$ we have
 $S_{\mu}(f_{t_0}):=\int f_{t_0}\log f_{t_0}\rd\mu \le CN^m$ with some fixed exponent $m$
that may depend on $\e$.
Let $E\in \bR$ and $b>0$ such that $\min\{\varrho(x)\; : \; x\in[E-b,E+b]\}>0$.
Then for any $\delta>0$, $\e'>0$,  
for any integer $n \ge 1$ and for any compactly supported continuous test function
$O:\bR^n\to \bR$, we have,
\be
\begin{split}
&  \sup_{t\ge \tau} \int_{E-b}^{E+b}\frac{\rd E'}{2b}
\int_{\R^n}   \rd\alpha_1
\ldots \rd\alpha_n \; O(\alpha_1,\ldots,\alpha_n) 
 \frac{1}{\varrho(E)^n}   \\
 & \quad \times \Big ( p_{t,N}^{(n)}  - p_{\mu, N} ^{(n)} \Big )
\Big (E'+\frac{\alpha_1}{N\varrho(E)},
\ldots, E'+\frac{\alpha_n}{ N\varrho(E)}\Big) 
\leq 
C N^{2 \e'} \Big[ b^{-1}  N^{ - \frac { 1 +  2 \e} 3 } +  b^{-1/2} N^{-\delta/2} \Big]
\label{abstrthm}
\end{split}
\ee
for $\tau = N^{-2 \e + \delta}$  where $\e>0$ is the exponent from Assumption III.

Suppose in addition to the Assumption I-IV,  that there exists an $A>0$ such that, for any $c\,'>0$
\be\label{memory}
\P \Big (  \sup_{c\,'N\leq j\leq  (1-c')N} |x_j-\gamma_j| 
\ge N^{-1+A} \Big ) \le C  N^{-c\log \log N }
\ee
for some constants $c$ and $C$ only depending on $c\,'$. 
Then for $\tau = N^{-2 \e + \delta}$   we have 
\be
\begin{split}
&  \sup_{t\ge \tau} \int_{E-b}^{E+b}\frac{\rd E'}{2b}
\int_{\R^n}   \rd\alpha_1
\ldots \rd\alpha_n \; O(\alpha_1,\ldots,\alpha_n) 
 \frac{1}{\varrho(E)^n} \\
& \quad \times   \Big ( p_{t,N}^{(n)}  - p_{\mu, N} ^{(n)} \Big )
\Big (E'+\frac{\alpha_1}{N\varrho(E)},
\ldots, E'+\frac{\alpha_k}{ N\varrho(E)}\Big)
\leq 
C_n N^{2 \e'} \Big[ b^{-1}  N^{ - 1+A } +  b^{-1/2} N^{-\delta/2} \Big].
\label{abstrthm2}
\end{split}
\ee

\end{theorem}

This theorem shows that the local  statistics of the points $x_j$ in the bulk
with respect to the time evolved distribution $f_t$ coincides with the local statistics
with respect to the equilibrium
distribution $\mu$ as long as $t\gg N^{-2\e}$. In many applications, the local equilibrium
statistics can be explicitly computed and in  the $b\to 0$ limit
it becomes independent of $E$, in particular this is the case for
the classical matrix ensembles (see next section).
 The restriction on
the time $t\gg N^{-2\e}$ will be removed by the reverse heat flow argument
(see Section \ref{sec:rev}) for matrix ensembles.

Since the eigenvalues fluctuate at least on a scale $1/N$, 
the best possible exponent in Assumption III is $2\e \sim 1$,
but  we will only be able  to prove it for some  $\e > 0$ for the ensembles considered 
 in this paper.
Similarly, the optimal exponent in \eqref{memory} is $A\sim 0$.
If we use these optimal estimates, $2\e\sim 1$, $A\sim 0$,
and we choose $\delta = 2\e \sim 1$, thus $\tau \sim 1$, 
then we can choose $b \sim N^{-1}$, i.e., we obtain
the universality with essentially no averaging in $E$.
On the other hand, the error estimate is the strongest,
of order  $\sim N^{-1/2}$, for an averaging on an energy
window of size $b\sim 1$. These errors become weaker if
time $\tau$ is reduced.
These considerations
are not important in this paper, but will be useful when good estimates on 
$\e$ and $A$ can be obtained.

\bigskip

{\it Convention:} Throughout the paper the letters $C, c$ denote positive
constants whose values may change from line to line and they
are independent of the relevant parameters. Since we will always take the $N\to \infty$
limit at the end, all estimates are understood for sufficiently large $N$.

\section{Universality for Matrix Ensembles}\label{sec:matr}

Now we specialize  Theorem \ref{thm:main} to Wigner and sample covariance matrices
with i.i.d. entries.  In the next sections we give the precise definitions
of these ensembles; formulas for the equilibrium measure and 
the dynamics will be deferred until
Section  \ref{sec:DBM}.

In order to apply Theorem \ref{thm:main} to Wigner and sample covariance matrices, 
we need to check that Assumptions I-IV are satisfied for these ensembles.
 Assumptions I or I' are satisfied by the definition of the Hamiltonian, 
the precise formulas are given in Section \ref{sec:DBM}. Assumption II is 
 satisfied since  the density
of eigenvalues is given by the Wigner semicircle law \eqref{def:rhosc}
for Wigner matrices \cite{W}. In case of the sample covariance
matrices, the singular values of $A$ will play the role of $x_j$'s
and their density is given by the Marchenko-Pastur law \eqref{defrhow}
after an obvious transformation \eqref{def:singv} \cite{MP}.
In fact, in Section \ref{sec:wish} we prove a local version
of the Marchenko-Pastur law in analogy with our
previous work on the local semicircle law for Wigner
matrices \cite{ESY1, ESY2}.
In Section \ref{sec:check} (Theorem  \ref{maintheoremWish}) we
will show that Assumption III is satisfied for these ensembles 
(more precisely, we will prove that Assumption III is satisfied for sample covariance matrices; 
the proof for Wigner matrices is analogous, and will not be given in details).
Assumption IV will be proved in Lemma  \ref{lm:up} for the sample covariance matrices,
for Wigner matrices the proof was given, e.g., in Theorem  4.6 of \cite{ESY3}.
 We remark that the assumption that the matrix entries are identically distributed, 
will only be used in checking Assumptions III and IV. Assumption II holds under
 much more general conditions on the matrix entries.
Finally, the apriori estimate on the entropy $S_\mu(f_{t_0})$ 
follows from the
smoothing property of the OU-flow (see Section \ref{sec:DBM}).
\bigskip

\subsection{Definition of the Wigner matrix}\label{sec:wigner}

To fix the notation, we assume that in the case of {\it real symmetric} matrices,
the matrix elements  of $H$ are given by
\be
 h_{\ell k} = h_{k\ell} : =N^{-1/2}    x_{\ell k}, \quad k <\ell,
\label{scaling}
\ee
where $x_{\ell k}$ for $\ell<k$ are independent,
identically distributed real random variables
with distribution $\nu$ that has zero expectation and
variance $1$.  The diagonal
elements  are $h_{kk} = N^{-1/2}x_{kk}$, where $x_{kk}$
are also i.i.d.  with distribution $\wt \nu$
that has zero expectation and   variance 2.
The eigenvalues   of $H$ will be denoted by $x_1 < x_2 <
\ldots < x_N$. We will always assume that the distribution
$\nu$ is continuous hence the eigenvalues are simple
with probability one.

In the {\it hermitian} case we assume that
\be
 h_{\ell k} =\bar h_{k\ell}:= N^{-1/2} (x_{\ell k}+iy_{\ell k}), \quad k <\ell,
\label{herm}
\ee
where $x_{\ell k}$ and $y_{\ell k}$ are real i.i.d.
random variables distributed with the  law $\nu$ with zero expectation and variance $\frac{1}{2}$.
The diagonal elements $h_{kk}$ are real, centered  and they have variance one with law $\wt\nu$.
The eigenvalues of $H$ are again denoted by $x_1 < x_2 <
\ldots < x_N$.

Finally, for the {\it quaternion self-dual} case we
assume that $H$ is a $2N$ by $2N$ complex matrix that
can be viewed as an $N\times N$ matrix  with elements consisting of $2\times 2$ blocks
of the form
\be
 \begin{pmatrix} z & w \\ -\bar w & \bar z \end{pmatrix},
\label{quatm}
\ee
where $z=a+bi, w=c+di$ are arbitrary complex numbers, $a,b,c,d\in\bR$. Such a 2 by 2 matrix can
be identified with the quaternion $q = a+ b{\bf i} + c{\bf j} + d{\bf k}\in \bH$
if the quaternion basis elements ${\bf i}, {\bf j}, {\bf k}$ are
identified with the standard Pauli matrices
$$
 {\bf i} = i\sigma_3= \begin{pmatrix} i & 0 \\ 0 & -i \end{pmatrix},\qquad
 {\bf j} = i\sigma_2= \begin{pmatrix} 0 & 1 \\ -1 & 0 \end{pmatrix},\qquad
{\bf k} = i\sigma_1= \begin{pmatrix} 0 & i \\ i & 0 \end{pmatrix}.
$$
The complex numbers $z\in \C$ can be naturally identified with
diagonal quaternions via the identification
\be
   z\cong \begin{pmatrix} z & 0 \\ 0 & \bar z \end{pmatrix}.
\label{idd}
\ee
The dual of the quaternion $q$ is defined to be $q^+:=
a- b{\bf i} - c{\bf j} - d{\bf k}$ which corresponds to the
hermitian conjugate of the matrix \eqref{quatm}.

Using this identification,
$H$ can be viewed as an $N\times N$ matrix with quaternion entries.
The matrix $H$ is quaternion self-dual if its entries
satisfy $ h_{\ell k} =h_{k\ell}^+$, in particular, the diagonal elements $h_{kk}$ are real.
We assume that the offdiagonal elements of $H$ are given (in the quaternion notation) by
\be
 h_{\ell k} =h_{k\ell}^+:
= N^{-1/2} (x_{\ell k}+{\bf i}y_{\ell k}+{\bf j} z_{\ell k} + {\bf k}u_{\ell k}), \quad 1\le
k < \ell\le N
\label{quat}
\ee
where $x_{\ell k}$, $y_{\ell k}$, $z_{\ell k}$ and $u_{\ell k}$ are real i.i.d.
random variables with  law $\nu$ that  has zero expectation and  variance $\frac{1}{4}$.
The diagonal entries are real,
$$
  h_{kk} = N^{-1/2} x_{kk}, \qquad 1\le k\le N,
$$
where $x_{kk}$ has a law $\wt \nu$ with zero expectation and variance $\frac{1}{2}$.
The spectrum of $H$ is doubly degenerate and we will neglect this degeneracy, i.e.,
we consider only  $N$ real (typically distinct) eigenvalues, $x_1<x_2<\ldots < x_N$.

The Gaussian ensembles (GOE, GUE and GSE) are
special Wigner ensembles with $\nu$ and $\wt\nu$ being Gaussian distribution.
These ensembles are invariant under their corresponding symmetry group,
i.e., the distribution remains unchanged under the conjugation
$H\to UHU^*$. Here $U$ is an arbitrary orthogonal matrix in case
of GOE, it is a unitary matrix for GUE and it is a unitary matrix over
the quaternions in case of GSE. In the latter case, if one uses
the $(2N)\times (2N)$ complex matrix representation, then the
symmetry group  is $\mbox{Sp}(N) = \mbox{Sp}(N,\C)\cap SU(2N)$.

With the given normalization, the eigenvalues are supported asymptotically
in $[-2,2]$ in all three cases,
Moreover their
empirical density converges weakly to the Wigner semicircle law in probability \cite{W},
i.e.,
for any $J\in C_0(\bR)$ and for any $\e>0$, we have
\be
\lim_{N\to\infty}
 \P\Bigg\{ \Big| \frac{1}{N}\sum_{j=1}^N J(x_j)  -
\int J(x)\varrho_{sc}(x)\rd x \Big|\ge \e\Bigg\} =0,
\label{wign}
\ee
where
\be
 \varrho_{sc}(x) : =\frac{1}{2\pi}\sqrt{(4-x^2)_+} \, .
\label{def:rhosc}
\ee
In particular,  the typical spacing between neighboring eigenvalues is of
order $1/N$ in the bulk of the spectrum.

We will often need to assume that the distributions
$\nu$ and $\wt\nu$ have Gaussian decay, i.e., there exists
$\delta_0>0$ such that
\be
    \int_\R \exp \big[ \delta_0 x^2\big] \rd\nu(x) < \infty,
\qquad \int_\R \exp \big[ \delta_0 x^2\big] \rd\wt\nu(x) < \infty.
\label{gauss}
\ee
In several statements we can relax this condition
to assuming only subexponential decay, i.e., that
there exists $\delta_0>0$ and $\gamma>0$ such that
\be
  \int e^{\delta_0 |x|^\gamma} \rd\nu(x) <\infty, \qquad
\int e^{\delta_0 |x|^\gamma} \rd\wt\nu(x) <\infty.
\label{subexp}
\ee
For some statements we will need to assume that
the measures $\nu$, $\wt\nu$ satisfy the logarithmic Sobolev inequality,
i.e., for
any density $h\ge 0$ with $ \int h \rd\nu =1$ it holds that
\be
 \int h\log h \, \rd\nu \le C\int |\nabla \sqrt{h}|^2\rd \nu
\label{LSI}
\ee
and a similar bound holds for $\wt\nu$.
We remark that \eqref{LSI} implies \eqref{gauss}, see, e.g. \cite{Le}.

\subsection{Sample Covariance Matrix}\label{sec:wishart}

The {\it real sample covariance} matrix ensemble consists of
symmetric $N\times N$  matrices of the form  $H=A^*A$. Here $A$ is an $M\times N$ real
matrix with $d=N/M$ fixed and we assume that $0<d<1$.
The elements of $A$ are given by
\be
 A_{\ell k} =   M^{-1/2} x_{\ell k},\,\,\, 1\leq \ell\leq M,\,\,\, 1\leq k\leq N,
\label{Wscaling}
\ee
where $x_{\ell k}$  are real i.i.d random variables
with the distribution $\nu$ that is symmetric and
has variance $1$. In the case of {\it complex sample covariance} ensemble we assume that
\be
 A_{\ell k} =   M^{-1/2}\left(  x_{\ell k}+iy_{\ell k}\right),\,\,\, 1\leq
\ell\leq M,\,\,\, 1\leq k\leq N,
\ee
where $x_{\ell k}$ and $y_{\ell k}$ are symmetric, real i.i.d.
random variables with  distribution $\nu$ that has variance $\frac{1}{2}$.
We will assume that $\nu$ has Gaussian  \eqref{gauss}
or sometimes only subexponential \eqref{subexp} decay.
The spectrum of $H$  asymptotically lies in the interval
$[\lambda_-, \lambda_+]$, where
\be\label{deflambda-+}
\lambda_\pm\equiv \left(1\pm d^{1/2}\right)^2.
\ee
Moreover, analogously to \eqref{wign}, the empirical density of eigenvalues converges weakly in
probability to the Marchenko-Pastur law
\be\label{defrhow}
\rho_W(x)=\frac{1}{2\pi d}\sqrt{\frac{\big[(\lambda_+-x)(x-\lambda_-)\big]_+}{x^2}} \; .
\ee
Most of the analysis will be done for the singular values of $A$ that
are denoted by $\bx=(x_1, \ldots , x_N)$. They are supported
asymptotically in $[\sqrt{\lambda_-}, \sqrt{\lambda_+}]$ and
therefore the typical spacing between neighboring singular values is of
order $ 1/N$. Their empirical density converges to
\be\label{def:singv}
 \wt \varrho_W(x): = 2x\varrho_W(x^2) =
\frac{1}{\pi d}\sqrt{\frac{\big[(\lambda_+-x^2)(x^2-\lambda_-)\big]_+}{x^2}}\; .
\ee

We remark that the  assumption that $\nu$ is
symmetric is used only at one technical step,  namely
when we refer to the large deviation result
for the extreme eigenvalues of the sample covariance matrices in \cite{FSo}
(see Lemma \ref{boundKK} below). The similar result for  Wigner
matrices has been proven without the symmetry condition,
see Theorem 1.4 in \cite{Vu}.

\subsection{Main Theorems}

With the remarks at the beginning of Section \ref{sec:matr},
Theorem \ref{thm:main} applies directly to prove  universality
for Wigner and sample covariance ensembles with a small Gaussian component;
we will not state these theorems separately.
To remove the small  time restriction from Theorem \ref{thm:main}, we
will apply the reverse heat flow argument.
This will give our main result:
\begin{theorem}\label{thm:main1} Consider an $N\times N$ symmetric, hermitian
or quaternion self-dual
Wigner  matrix $H$, or an $N\times N$ real or complex sample covariance matrix $A^*A$. Assume
that the single site entries of $H$ or $A$  are i.i.d. with
probability distribution $\nu(\rd x)= u_0(x) \rd x$ and with the standard
normalization specified in Sections \ref{sec:wigner} and \ref{sec:wishart}.
We assume that $\nu$ satisfies the logarithmic Sobolev inequality \eqref{LSI}
and in case of the sample covariance matrix we also assume that $\nu$ is symmetric.
The same conditions are assumed for the distribution $\wt \nu$ of the diagonal
elements in case of the Wigner
matrix. Let $f_0=f_{0,N}$ denote the joint density
function of the eigenvalues and
let $p_{0,N}^{(k)}$ be the $k$-point correlation
function of $f_0$.  Let $\varrho$ denote the corresponding
density of states, i.e.,
$\varrho$ is given by the Wigner semicircle law \eqref{def:rhosc} or the Marchenko-Pastur law
\eqref{defrhow}, respectively.
Let $E\in \bR$, $b>0$  such that $\min\{\varrho(x)\; : \; x\in[E-b,E+b]\}>0$. If
for any $k\ge 1$ there is a constants $M_k$ 
such that  the density function $u_0$ satisfies
\be
  \sum_{j=0}^{M_k} |\pt_x^j\log u_0(x)| \le C_k(1+|x|)^{C_k}
\label{smooth}
\ee
for some constants $C_k < \infty$, then for any compactly supported continuous test function
$O:\bR^k\to \bR$ we have
\be
\begin{split}
\lim_{N\to \infty}
\int_{E-b}^{E+b}\rd E' \int_{\R^k} & \rd\alpha_1
\ldots \rd\alpha_k \; O(\alpha_1,\ldots,\alpha_k)  \\
&\times
\frac{1}{\varrho(E)^k} \Big ( p_{0,N}^{(k)}  - p_{\mu, N} ^{(k)} \Big )
\Big (E'+\frac{\alpha_1}{N\varrho(E)},
\ldots, E'+\frac{\alpha_k}{N \varrho(E)}\Big) =0.
\label{matrixthm}
\end{split}
\ee
Here $\mu$ denotes the probability measure of the eigenvalues of the
 appropriate Gaussian ensemble,
i.e. GUE, GOE, GSE for the case of hermitian, symmetric, and,
 respectively, quaternion self-dual Wigner
 matrices;  and the ensembles of real or complex sample covariance matrices with Gaussian entries
(Wishart ensemble) in case of the  covariance matrices $A^*A$.
These measures are given in (\ref{Huj}), with $\beta=1,2,4$,
for Wigner matrices, and, expressed in terms of singular values, in (\ref{HW}),
 with $\beta=1,2$, for sample covariance matrices.
\end{theorem}

{\bf Remark 1.:}
In the case of symmetric and hermitian Wigner matrices, the condition
\eqref{smooth} can be removed by
applying the Four-moment theorem of Tao and Vu (Theorem 15 of \cite{TV})
as in the proof of Corollary 2.4 of \cite{ESY4}.
Similar remark applies to the sample covariance ensembles and to the
quaternion self-dual  Wigner ensemble provided the corresponding
Four-moment theorem is established.

We also remark that a manuscript by Ben-Arous and P\'ech\'e
\cite{BP1} with a similar statement is in preparation
for complex sample covariance matrices that holds for a fixed $E'$, i.e.,
without averaging over the energy parameter in  \eqref{matrixthm}.

{\bf Remark 2.:} 
After the first version
of this manuscript was posted on the arxiv, the question that whether  the four
moment theorem for  sample covariance matrices holds
was settled  in \cite{TV3}. In particular, \cite{TV3}  gives an alternative
proof of the universality  of local statistics
for the complex  sample covariance ensemble when
combined with the result of \cite{BP}. 
For the real sample covariance ensemble the universality was established
for  distributions  whose  first four moments
match the standard  Gaussian variable.   An important common ingredient
to both our approach and that of \cite{TV3} 
is  the  local Marchenko-Pastur law, 
established  in  Proposition \ref{mainprop}; a slightly different version 
 suitable for the  application to prove the four moment theorem is 
proved  in \cite{TV3}. 

The four moment theorem in \cite{TV3} compares the distributions of 
individual eigenvalues
for two different ensembles. For our application to the correlation
 functions and gap distributions, 
an alternative approach is to use the recent Green function comparison
 theorem \cite{EYY}. This will also remove 
the smoothness and logarithmic Sobolev inequality restrictions in
  Theorem \ref{thm:main1}.

\bigskip

We now state our result concerning  the eigenvalue gap distribution
both for Wigner and sample covariance ensembles.
For any $s>0$ and $E$ with $\rho(E) > 0$  we
define the density  of eigenvalue pairs  with distance less than
$s/N\varrho(E)$ in the vicinity of $E$ by
\be
\Lambda (E; s) =
\frac {1}{2  N \ell_N \varrho (E)}
\# \Big\{ 1\le j \le N-1\,: \; x_{j+1} - x_j \le
\frac{s}{N\varrho (E)}, \; |x_j-E| \le \ell_N  \Big\},
\label{def:Lambda}
\ee
where $\ell_N = N^{- \delta}$ for some $0< \delta\ll  1$.

\begin{theorem}\label{mainthm2}
Consider an $N\times N$ Wigner or sample covariance matrix as in
Theorem \ref{thm:main1} such  that
the probability measure $\rd\nu=u_0\rd x$ of the matrix elements
satisfies the logarithmic Sobolev inequality \eqref{LSI}
and, additionally, $\nu$ is symmetric in the sample covariance matrix case.
Suppose that the initial  density $u_0$ satisfies
\be
  \sum_{j=0}^M |\pt_x^j\log u_0(x)| \le C(1+|x|)^C
\label{smooth1}
\ee
with some sufficiently large constants $C, M$ that depend on the $\e$ in Assumption III.
Then
for any $E$ with $\rho(E) > 0$  and for any continuous, compactly supported
test function $O:\bR\to\bR$ we have
\be\label{maineq2}
\lim_{N\to \infty}   \int_\bR \rd s  \, O(s)  
 [  \E \, \Lambda (E; s) - \E^\mu  \, \Lambda (E; s)]
=0, 
\ee
where $\mu$ is the probability measure of the eigenvalues of the
 appropriate Gaussian ensemble, as in Theorem \ref{thm:main1}.
\end{theorem}

Theorem \ref{mainthm2} shows that, in particular, the
probability to find no eigenvalue in the
interval $[E, E+\al/(\varrho (E) N)]$   is asymptotically the same 
as in the corresponding classical Gaussian ensemble. 
Theorems \ref{thm:main1} and \ref{mainthm2}
will follow from Theorem \ref{thm:main} and the  reverse heat flow argument
that we present in  Section \ref{sec:rev}.
We remark that the additional condition on the symmetry
of $\nu$ in the case of sample covariance matrices stems from
using a result from  \cite{FSo} on the lowest eigenvalue
of these matrices, see Lemma \ref{boundKK}.

Theorem \ref{mainthm2} can be proven directly
from  Theorem \ref{thmM}  since the test functions of the form
\[
 \frac 1 N \sum_{i\in J} G(N(x_i-x_{i+1}))
\]
determine the distribution of the random variable $\Lambda (E; s)$ uniquely.
Here we take $J$ to be the set
$$
  J : = \big\{ i\; : \; \gamma_i \in [E-\ell_N, E+\ell_N]\big\},\quad
$$
where $\gamma_i$ was defined in \eqref{gammaj}.
Notice  that $\delta$ in the definition of $\ell_N$ has to be  small enough
so that the edge term near the boundary of the interval is negligible.

\section{Local Relaxation Flow}

\begin{theorem}[Universality of Dyson Brownian Motion for Short Time] \label{thmM}
Suppose  that the  Hamiltonian $\cH$ given in \eqref{ham} satisfies
the convexity bound \eqref{convex} with $\beta\ge1$. Let $f_t$ be the solution
of the forward equation \eqref{dy} with an initial density $f_0$.
Fix a positive   $\e>0$, set $t_0=N^{-2\e}$ and define
\be\label{Q}
  Q:= \sup_{t\ge t_0}    \sum_j \int ( x_j-\gamma_j)^2 f_t \rd \mu .
\ee
Assume that at time $t_0$ we have
 $S_{\mu}(f_{t_0}):=\int f_{t_0}\log f_{t_0}\rd\mu \le CN^m$ with some fixed exponent $m$
that may depend on $\e$. 
Fix $n\ge 1$ and an array of positive  integers, $\bm = (m_1, m_2, \ldots, m_n)\in
\N^n_+$.
Let  $G:\bR^n\to\bR$ be a bounded smooth function with compact 
support and define
\be
  \cG_{i,\bm}(\bx) := 
G\Big( N(x_i-x_{i+m_1}), N(x_{i+m_1}-x_{i+m_2}), \ldots, N(x_{i+m_{n-1}}-x_{i+m_n})\Big).
\label{cG}
\ee
Then  for any sufficiently small $\e'>0$,
there exist
constants $C, c$, depending only on $\e'$ and
 $G$ such that
 for any $J\subset \{ 1, 2, \ldots , N-m_n\}$ and for  any $\tau \ge 3t_0=3N^{-2\e}$,  we have
\be\label{GG}
\Big| \int \frac 1 N \sum_{i\in J} \cG_{i,\bm}(\bx) f_\tau \rd \mu -
\int \frac 1 N \sum_{i\in J} \cG_{i,\bm}(\bx) \rd\mu \Big|
\le C N^{\e'}  \sqrt{|J|Q(\tau N)^{-1}}  + Ce^{-cN^{\e'}} ,
\ee
where $|J|$ is the number of the elements in $J$. 
\end{theorem}

The proof of this theorem is similar but
much simpler than  that of
Theorem 2.1 of \cite{ESY4}.  
The estimate \eqref{GG}  improves slightly over the similar estimate in  \cite{ESY4}
by a   factor $|J|/N$  due to  the  improvement in \eqref{4.1}.
Theorem \ref{thm:main} will  follow
from the fact that in case $\tau\ge N^{-2\e+\delta}$, the assumption
\eqref{assum2} guarantees that
$$
N^{\e'}\sqrt{|J|Q(\tau N)^{-1}} \le  N^{\e'-\delta/2} = N^{-\delta/6}\to 0
$$
with the choice $\e'=\delta/3$ and using $|J|\le N$.
More precise error bound will be obtained by relating $b$ to $|J|$.
 Therefore the local statistics
of observables involving eigenvalue differences coincide
in the $N\to\infty$ limit.  To complete the proof of  Theorem \ref{thm:main}, we will
have to show that the convergence of the
 observables  $\cG_{i,\bm}$ is sufficient to identify
the correlation functions of the $x_i$'s in the sense prescribed in 
 Theorem \ref{thm:main}. The details will be given in 
Section \ref{sec:corfn}.

\bigskip

{\it Proof of Theorem \ref{thmM}.} Without loss of generality
we can assume in the sequel that $f_0\in L^\infty(\rd\mu)$. 
To see this, note that
any $f_0\in L^1(\rd\mu)$ can be  approximated
by a sequence of  bounded functions $f_0^{(k)}$ in $L^1$-norm with
arbitrary precision and the dynamics is a contraction
in $L^1$ (see Appendix \ref{sec:defdyn}), thus $f_\tau$
and $f_\tau^{(k)}$ are arbitrarily close in $L^1$.
Since $G$ is bounded on the left hand side of \eqref{GG},
this is sufficient to pass to the limit $k\to\infty$.

Every constant in this proof depends on $\e'$ and $G$, and
we will not follow the precise dependence. We can assume that $\e'<\e$.
Given $\tau >0$, we define 
\be
R:= \tau^{1/2} N^{-\e'/2}.
\ee
Notice that the choice of $R$ depending on $\tau$ which is the main reason that $\tau$ appears 
in the denominator on the right hand side of  \eqref{GG}.
We now introduce the {\it pseudo equilibrium  measure}, $\om_N=\om= \psi\mu_N$,
defined by
\[
\psi=\frac{Z}{\wt Z}\exp \big(-NW \big),  \quad W(\bx) =    \sum_{j=1}^N
W_j (x_j)   , \qquad W_j (x) = \frac{1}{2R^2} (x_j -\gamma_j)^2,
\]
where $\wt Z$ is chosen such that $\om$ is a probability measure,
in particular $\omega= e^{-N\wt \cH}/\wt Z$ with
\be
\wt \cH = \cH +W.
\label{def:wth}
\ee
Similarly to \eqref{Zbound}, one can check that 
\be
 |\log \wt Z|\le CN^m
\label{Ztildebound}
\ee
with some exponent $m$.

Note that the additional term $W_j$ confines
the $j$-th point $x_j$ near its classical location. We will prove that the probability w.r.t. the 
equilibrium measure $\mu_N$ of the event that 
$x_j$ near its classical location is very close to $1$. Thus there is little difference between 
the two measures $\om_N$ and $\mu_N$ and in fact, we will prove that their local statistics 
are identical in the limit $N \to \infty$.  The   main 
advantage of the pseudo equilibrium  measure comes from the fact
that it has a faster decay to global equilibrium  
as shown in Theorem \ref{thm2}.

The local relaxation flow is defined to be  the reversible dynamics w.r.t.
$\omega$.
The dynamics is described by the generator  $\tilde L$ defined by
\be\label{Lt}
\int f \tilde L  g \rd \omega = - \frac 1 {2N} \sum_j
\int \partial_j f \partial_j g \rd \omega .
\ee
Explicitly,  $\wt L$ is given by
\be\label{tl}
L = \wt L + \sum_j b_j \partial_j, \quad
b_j = W_j'(x_j)= \frac{x_j -\gamma_j}{R^2}.
\ee
Since the additional potential $W_j$ is uniformly convex with
\be\label{5.9}
\inf_j \inf_{ x \in \bR}     W_j^{\prime \prime}(x)
\ge R^{-2},
\ee
by \eqref{convex} and $\beta \ge 1$ we have
\be\label{convex1}
\Big\langle \bv , \nabla^2 \wt\cH(\bx)\bv\Big\rangle
\ge   \frac{1}{R^2} \,  \|\bv\|^2 + \frac{1}{N}
 \sum_{i<j} \frac{(v_i - v_j)^2}{(x_i-x_j)^2}, \qquad \bv\in\bR^N.
\ee
Here we have used $U''\ge 0$ in the last estimate. If this assumption is replaced by 
\be\label{UprimeM}
U^{\prime \prime} \ge - M
\ee
for some constant $M$ independent of $N$, then there will be an extra term $- M\|\bv\|^2$ 
in \eqref{convex1}. Assuming $\tau \le N^{\e'}$, we have $R\le N^{-\e'/2}$, then
this extra term can be controlled by the  $R^{-2}$ term and the same proof will go through. 
Since for the applications in this paper, the condition $ U^{\prime \prime} \ge 0$ is satisfied, 
we will not use this remark here.

The $R^{-2}$ in the first term comes from the additional convexity of
the local interaction and it enhances the  ``local
Dirichlet form dissipation''.
In particular we have the uniform lower bound
\be\label{hess}
\nabla^2\wt \cH =\mbox{Hess}  (- \log \omega) \ge   R^{-2}.
\ee
This guarantees that the relaxation time to equilibrium $\om$
for the $\wt L$ dynamics is bounded above by  $C R^2$.
We recall the definition of the relative entropy of $f$
with respect to any probability measure $\rd\lambda$
\[
S_\lambda(f) = \int f\log f \rd \lambda, \qquad
S_\lambda (f|\psi)=  \int f \log (f/\psi) \rd \lambda.
\]

\bigskip

The first ingredient to prove Theorem \ref{thmM} is the analysis of
the local relaxation flow which  satisfies the logarithmic
Sobolev inequality and the following
dissipation estimate. Its  proof  follows the standard argument in \cite{BE} (used
in this context in Section 5.1 of \cite{ERSY}).  In Appendix \ref{sec:BE} we will explain how
to extend this argument onto the subdomain $\Sigma_N$.
Here we only remark that  the key inputs are the convexity 
bounds (\ref{convex1}, \ref{hess}) on the Hessian  of $\wt \cH$ \eqref{convex1}.

\begin{theorem}\label{thm2}
Suppose \eqref{convex1}  holds.
Consider the forward equation
\be
\partial_t q_t=\wt L q_t, \qquad t\ge 0,
\label{dytilde}
\ee
with an initial condition $q_0$ and
with the reversible measure $\omega$. Assume that $q_0\in L^\infty(\rd\om)$.
Then we have the following estimates
\be\label{0.1}
\partial_t D_{\omega}( \sqrt {q_t}) \le - \frac{1}{R^2}D_{\omega}( \sqrt {q_t}) -
\frac{1}{2N^2}  \int    \sum_{i,j=1}^N
\frac{  ( \pt_i \sqrt{ q_t} - \pt_j\sqrt {q_t} )^2}{(x_i-x_j)^2} \rd \omega ,
\ee
\be\label{0.2}
\frac{1}{2N^2} \int_0^\infty  \rd s  \int    \sum_{ i,j=1}^N
\frac{(\pt_i\sqrt {q_s} - \pt_j\sqrt {q_s} )^2 }{(x_i-x_j)^2}\rd \omega
\le D_{\omega}( \sqrt {q_0})
\ee
and the logarithmic Sobolev inequality
\be\label{lsi}
S_{\omega}(q)\le C R^2  D_{\omega}( \sqrt {q})
\ee
with a universal constant $C$.
Thus the time to equilibrium is of order $R^2$:
\be\label{Sdecay}
 S_{\omega}(q_t)\le e^{-Ct/R^2} S_\omega(q_0).
\ee
\qed
\end{theorem}

The estimate \eqref{0.2} on the second  term in \eqref{convex1}
plays a key role in the next theorem.

\begin{theorem}\label{thm3}
Suppose that Assumption I holds and we have  
a density  $q\in L^\infty$, $\int q\rd\om =1$.
Recall that $\tau = R^2N^{\e'}$.
Fix $n\ge 1$, $\bm\in \cN_+^n$, 
let  $G:\bR^n\to\bR$ be a bounded smooth function with compact support
and recall the definition of $\cG_{i,\bm}$ from \eqref{cG}.
Then  for any $J\subset \{ 1, 2, \ldots , N-n\}$  we have
\be\label{diff}
\Big| \int \frac 1 N \sum_{i\in J} \cG_{i,\bm}(\bx) \rd \omega -
\int \frac 1 N \sum_{i\in J} \cG_{i,\bm}(\bx) q \rd \omega \Big|
\le C \Big( \frac {|J|D_\omega (\sqrt {q})  \tau}{N^2}  \Big)^{1/2}
 + C e^{-cN^{\e'}} \sqrt{S_\om(q)}.
\ee
\end{theorem}

{\it Proof.} For simplicity, we will consider the case when $\bm =(1,2,\ldots n)$,
the general case easily follows by appropriately redefining the  function $G$.
Let $q_t$ satisfy
\[
\partial_t q_t = \wt L q_t, \qquad t\ge 0,
\]
with an initial condition $q$.
Thanks to the exponential decay of the entropy on time
scale $\tau\gg R^2$, see \eqref{Sdecay}, and the entropy bound on  the initial state $q$,
the difference between
the local statistics  w.r.t. $q_\tau \om $ and $q_\infty \om =\om$ is subexponentially
small in $N$,
\begin{align}
\Big| \int \frac 1 N \sum_{i\in J} \cG_{i,\bm}(\bx) q_\tau \rd \omega -
\int \frac 1 N \sum_{i\in J} \cG_{i,\bm}(\bx) q_\infty \rd \omega \Big|
 \le & \| G\|_\infty \int |q_\tau-1|\rd\om \non \\
 \le & C\sqrt {S_\om(q_\tau)} 
\le  C e^{-cN^{\e'}}\sqrt {S_\om(q)} , \non
\end{align}
 giving the second term on the r.h.s. of \eqref{diff}.
To compare $q$ with $q_\tau$,
by differentiation, we have
\begin{align}
\int \frac 1 N \sum_{i\in J} &\cG_{i,\bm}(\bx)q_\tau \rd \omega  -
\int \frac 1 N \sum_{i\in J} \cG_{i,\bm}(\bx)q \rd \omega \non\\
&= \int_0^\tau  \rd s \int  \frac 1 N \sum_{i\in J} \sum_{k=1}^n
   \pt_k G\Big( N(x_i-x_{i+1}), \ldots, N(x_{i+n-1}-x_{i+n})\Big)
[\pt_{i+k-1} q_s - \pt_{i+k}q_s]  \rd \omega. \non
\end{align}
Here we used the definition of $\wt L$ from \eqref{Lt} and note that
the $1/N$ factor present in \eqref{Lt} cancels the factor $N$
from the argument of $G$ \eqref{cG}.
 {F}rom the Schwarz inequality and $\pt q = 2 \sqrt{q}\pt\sqrt{q}$,
the last term is bounded by
\begin{align}\label{4.1}
2\sum_{k=1}^n & \left [   \int_0^\tau  \rd s \int
\sum_{i\in J}  \Big[\pt_k G \big(N(x_i - x_{i +1}),\ldots, N(x_{i+n-1}-x_{i+n}) \big)\Big] ^2 
(x_{i+k-1}-x_{i+k})^2  \, q_s \rd \omega
\right ]^{1/2} \nonumber \\
& \times \left [ \int_0^\tau  \rd s \int  \frac 1 {N^2 } \sum_{i\in J}
\frac{1}{(x_{i+k-1}-x_{i+k})^2}  [ \pt_{i+k-1}\sqrt {q_s} -
\pt_{i+k}\sqrt {q_s}]^2  \rd \omega \right ]^{1/2} \nonumber \\
\le &  \; C \sqrt { \frac{|J|D_\omega(\sqrt {q}) \tau}{N^2}},
\end{align}
where we have used \eqref{0.2} and that
$$
  \Big[\pt_k G \Big(N(x_i - x_{i +1}), \ldots, N(x_{i+k-1}-x_{i+k}), \ldots
 N(x_{i+n-1}-x_{i+n}) \Big) \Big]^2
 (x_{i+k-1}-x_{i+k})^2 \le CN^{-2},
$$
since  $G$ is smooth and  compactly
supported. This proves Theorem \ref{thm3}.
\qed

\bigskip

As a comparison to Theorem \ref{thm3}, we state the following result which can
be proved in a similar way.

\begin{lemma}\label{thm4}
Let  $G:\bR\to\bR$
 be a bounded smooth function with compact support and let
a sequence $E_i$ be fixed.
Then  we have
\be\label{Diff}
\Big|\frac 1 N \sum_{i}   \int G\big( N(x_i-E_i )\big) \rd \omega -
\frac 1 N \sum_{i}\int  G\big(N (x_i-E_i )\big)  q \rd \omega \Big|
\le C\Big( S_\omega (q)  \tau \Big)^{1/2}+  Ce^{-cN^{\e'}}\sqrt {S_\om(q)} .
\ee
\end{lemma}

Notice that  by exploiting the local Dirichlet form dissipation
coming from the second term on the r.h.s. of \eqref{0.1},
we have gained the crucial  factor $N^{-1/2}$
in the estimate \eqref{diff} compared with \eqref{Diff}.

\bigskip

The final  ingredient to prove Theorem \ref{thmM} is the following entropy and
Dirichlet form estimates.

\begin{theorem}\label{thm1}
Suppose that  \eqref{convex} holds and recall $\tau = R^2 N^{\e'}\ge 3t_0$ with $t_0=N^{-2\e}$.
Let  $g_t = f_t/\psi$ so that
$S_{\mu} (f_t|\psi) =  S_{\omega} (g_t)$.
Assume that $S_\mu(f_{t_0}) \le CN^m$ with some fixed $m$.
Then the entropy and the Dirichlet form satisfy the estimates:
\be\label{1.3}
S_{\omega} (g_{\tau/2}) \le
 C  NR^{-2}   Q, \qquad
D_\omega (\sqrt{g_\tau})
\le CN R^{-4} Q.
\ee
\end{theorem}

{\it Proof.}  Recall that  $\pt_t f_t = Lf_t$.
The standard estimate on the entropy of $f_t$ with respect
to the invariant measure  is obtained by differentiating
the entropy twice and using the logarithmic Sobolev inequality. The entropy
and the Dirichlet form in \eqref{1.3} are, however, computed
with respect to the measure $\om$. This yields the additional
second  term in the following identity  \cite{Y} that holds for
any probability density
$\psi_t$:
$$
\partial_t  S_\mu(f_t|\psi_t) = -  \frac{2}{N}  \sum_{j} \int (\partial_j
\sqrt {g_t})^2  \, \psi_t \, \rd\mu
+\int g_t(L-\partial_t)\psi_t \, \rd\mu \ ,
$$
where $g_t = f_t/\psi_t$. In our application we set $\psi_t$ to be time independent,
$\psi_t = \psi =\om/\mu$,
hence we have
$$
\pt_t S_\omega(g_t)
= - \frac{ 2}{N}  \sum_{j} \int (\partial_j
\sqrt {g_t})^2  \, \rd\omega
+\int   \wt L g_t  \, \rd\omega+  \sum_{j} \int  b_j \partial_j
g_t   \, \rd\omega.
$$
Since $\om$ is invariant, the middle term on the right hand side vanishes, and
from the Schwarz inequality
\be\label{1.1}
\pt_t S_\omega(g_t) \le  -D_{\omega} (\sqrt {g_t})
+   C  N\sum_{j} \int  b_j^2  g_t   \, \rd\omega \le
-D_{\omega} (\sqrt {g_t})
+   C  N\Lambda, \qquad t\ge N^{-2\e},
\ee
where we defined
\be\label{bbound}
\Lambda:= Q R^{-4} = \sup_{t\ge N^{-2\e}}  R^{-4}  \sum_j \int ( x_j-
\gamma_j)^2 f_t \rd \mu .
\ee
Together with  \eqref{lsi},  we have
\be\label{1.2new}
\partial_t  S_\omega(g_t) \le  - CR^{-2}  S_\omega(g_t) +
 C  N \Lambda, \qquad t\ge N^{-2\e}.
\ee
To obtain the first inequality in  \eqref{1.3},
we integrate \eqref{1.2new} from $t_0=N^{-2\e}$ to $\tau/2$, using that $\tau= R^{2} N^{\e'}$ and
$S_\om(g_{t_0})\le CN^m+N^2Q$ with some finite $m$, depending on $\e$. This apriori bound 
follows from 
\be
  S_\om(g_{t_0}) = S_\mu(f_{t_0}|\psi) =
 S_\mu(f_{t_0}) - \log Z + \log \wt Z + N \int f_{t_0} W\rd\mu
 \le CN^m + N^2Q,
\label{inentr}
\ee
where we used \eqref{Zbound} and \eqref{Ztildebound}.
The second inequality in \eqref{1.3} can be obtained from the first one
by integrating  \eqref{1.1}
from $t=\tau/2$ to $t=\tau$ and using
the monotonicity of the Dirichlet form in time. \qed

\bigskip

Finally, we complete the proof of Theorem \ref{thmM}.
Recall that 
$\tau = R^2 N^{\e'}$ and $t_0=N^{-2\e}$.
Choose $q_\tau:= g_{\tau}= f_{\tau}/\psi$ as density $q$ in
Theorem \ref{thm3}.
The condition $q_\tau\in L^\infty$ can
be guaranteed by the approximation argument from the beginning of
the proof of Theorem \ref{thmM}.
 Then  Theorem \ref{thm1}, Theorem \ref{thm3} together with \eqref{inentr} and the 
fact that $\Lambda \tau  = Q\tau^{-1}N^{2\e'}$
directly imply that 
\be\label{diff1}
\Big| \int \frac 1 N \sum_{i\in J} \cG_{i,\bm} f_\tau\rd \mu -
\int \frac 1 N \sum_{i\in J} \cG_{i,\bm}  \rd \omega \Big| \\
\le C N^{\e'}\sqrt{|J|Q (\tau N)^{-1} }
+  Ce^{-cN^{\e'}} ,
\ee
i.e., the local statistics of $f_\tau \mu$ and $\om$
can be compared.
Clearly, equation  \eqref{diff1}  also holds for the special choice
$f_0= 1$ (for which $f_\tau=1$), i.e., local statistics of $\mu$ and
$\om$ can also be compared. This completes the proof of Theorem \ref{thmM}.
\qed

\section{Equilibrium measure and Dyson Brownian motion}\label{sec:DBM}

We will treat the Wigner and sample covariance ensembles in parallel.
Suppose $(x_1, x_2, \ldots , x_N)$ denote the eigenvalues of
the Gaussian Wigner ensembles.
The joint distribution of $\bx= (x_1, x_2, \ldots , x_N)\in \bR^N$
of the Gaussian Wigner ensembles
is given by the following measure on $\bR^N$
\be\label{H}
\mu=\mu_{\beta, N}(\rd{\bf x})=
\frac{e^{-N\cH_\beta({\bf x})}}{Z_\beta}\rd{\bf x},\qquad \cH_\beta({\bf x}) =
\beta \left [ \sum_{i=1}^N \frac{1}{4} x_{i}^{2}  -  \frac{1}{N} \sum_{i< j}
\log |x_{j} - x_{i}| \right ]
\ee
where $\beta\ge 1$ is an arbitrary parameter, i.e., this corresponds to
choosing $U(x) =  x^2/4$ in \eqref{ham}.
With a slight abuse of notations we will use $\mu$ for both the measure
$\rd\mu$ and its density $e^{-N\cH_\beta}/Z_\beta$ with respect to the Lebesgue measure.
The specific
value $\beta =1, 2, 4$ correspond to the GUE, GOE and GSE ensembles, respectively.

We define the following generator
\be
L= L_{\beta,N}=  \sum_{i=1}^N \frac{1}{2N}\partial_{i}^{2} +\beta\sum_{i=1}^N
\Bigg(- \frac{1}{4} x_{i} +  \frac{1}{2N}\sum_{j\ne i}
\frac{1}{x_i - x_j}\Bigg) \partial_{i}
\label{L}
\ee
acting on $L^2(\mu)$. The measure $\mu$ is invariant
and reversible with respect to the dynamics generated by $L$.
Define the Dirichlet form and entropy by
\be
D(f) = D_\mu(f) =  -\int  f L f  \rd \mu =  \sum_{j=1}^N \frac{1}{2N}
\int (\partial_j f)^2 \rd \mu, \quad \mbox{and} \quad
S(f) = S_\mu(f) : =\int f\log f \rd\mu
\label{def:dir}
\ee

Let $f_t\rd\mu$ denote the probability measure  on the set
$\Sigma_N$ at the time $t$ with the given generator $L$.  Then $f_t$ satisfies the forward equation
\be
\partial_{t} f_t =  L f_t
\ee
with initial condition $f_0$. This dynamics is the Dyson Brownian motion.

The Dyson Brownian motion is the
corresponding system of stochastic differential equations
for the vector ${\bf x}(t)$ that is  given by
\be\label{sde}
\rd  x_i  =    \frac{\rd B_i}{\sqrt{N}} +  \beta \left [ -  \frac{1}{4}
x_i+  \frac{1}{2 N}\sum_{j\ne i}
\frac{1}{x_i - x_j}  \right ]  \rd t, \qquad 1\leq i\leq N,
\ee
where $\{ B_i\; : \; 1\leq i\leq N\}$ is a collection of
independent standard Brownian motions on $\bR$.  This SDE is well posed
for $\beta \ge 1$, and in particular the points do not cross each
other with probability one, i.e., the process is well defined on $\Sigma_N$
(see, e.g. Section 12.1 of \cite{G})

\bigskip

The treatment of the sample covariance ensembles is fully analogous, but the
formulas change slightly.  We use the
convention in the sample covariance case that $x_i$ denotes the singular values of $A$
and $\lambda_i = x_i^2$ are the eigenvalues of $A^*A$. Most of the formulas
will be in terms of $x_i$'s;  in particular we consider
the joint distribution function $f_0(\bx)$ of the singular values.
The invariant measure for the singular values is given by
(c.f. \eqref{H}):
\be\label{HW}
\mu^W=\mu_{\beta, N}^W(\rd{\bf x})=
\frac{e^{-N\cH_\beta^W({\bf x})}}{Z_\beta}\rd{\bf x}.
\ee
where
$$
\cH_\beta^W({\bf x}) =
\beta\left [ \sum_{i=1}^N \frac{x_{i}^{2}}{2d } -  \frac{1}{N} \sum_{i< j}
\log |x_{j}^2 - x_{i}^2| - \Big( \frac1d-1 +\frac{1-\beta^{-1}}{N}\Big)
\sum_{i=1}^N \log |x_i| \right ],
$$
where $d=M/N$  and  $\beta=1$ when $X$ is a real matrix, $\beta=2$ when $X$ is a complex matrix.
This formula can be obtained by direct calculation (see also
Proposition 2.16 of \cite{For} or Fig. 1 of \cite{DE} after appropriate rescaling).
Define the generator (c.f. \eqref{L})
\be
L^W= L_{\beta,N}^W=  \sum_{i=1}^N \frac{1}{2N}\partial_{i}^{2} +\sum_{i=1}^N
\Bigg(- \frac{\beta x_{i}}{2d}  +  \frac{\beta }{N}\sum_{j\ne i}
\frac{x_i}{x_i^2 - x_j^2}  +  \frac{1}{2}\Big(\beta\big(\frac1d-1\big) +\frac{\beta-1}{N}\Big)
\frac{1}{x_j}
\Bigg) \partial_{i}.
\label{LW2}
\ee
Finally, the stochastic differential equation is given by  (c.f. \eqref{sde})
\be\label{sdeW2}
\rd  x_i  =    \frac{\rd B_i}{\sqrt{N}} +  \left [ -   \frac{\beta x_{i}}{2d} +
\frac{\beta}{ 2N }\sum_{j\ne i} \left [
\frac{1}{x_i - x_j} +  \frac{1}{x_i + x_j} \right ] +
  \frac{1}{2}\Big(\beta\big(\frac{1}{d}-1\big) +\frac{\beta-1}{N}\Big)
\frac{1}{x_j}  \right ]  \rd t, \qquad 1\leq i\leq N.
\ee

\bigskip

In applications to Wigner matrices ($\beta =1,2, 4$), $f_0\rd\mu$
will be the joint probability density of the eigenvalues
of the initial hermitian, symmetric or quaternion self-dual Wigner matrix $\wh H$.
The limiting density is the Wigner semicircle law given in \eqref{def:rhosc}.
The Dyson Brownian motion describes the  eigenvalues
of the matrix valued process
\be
 \rd H_t = \frac{\rd {\bf B}_t}{\sqrt{N}} - \frac{1}{2} H_t \rd t
\label{Hsde}
\ee
with $H_0= \wh H$. Here ${\bf B}_t$ is a symmetric, hermitian or quaternion self-dual
matrix-valued process whose
offdiagonal elements are standard real, complex or quaternion Brownian
motions with variance one and
the diagonal elements of  are real Brownian motions with variance
$2,1$ and $\frac{1}{2}$, in case $\beta=1,2, 4$, respectively.
More precisely, let $u_t$ denote the density function of
the distribution of one real component of the $(ij)$-th entry of $H_t$, $i< j$
(there are two real components for the hermitian matrices and four
for the quaternion matrices), then
\be\label{ou}
\partial_{t} u_t =  B u_t, \quad
 B  = \frac{1}{2}\frac{\pt^2}{\pt x^2}
- \frac{\beta x}{2} \frac{\pt}{\pt x}.
\ee
Let  $\gamma(\rd x) = \gamma(x)\rd x:=(\beta/2\pi)^{1/2}e^{-\beta x^2/2} \rd x$
denote the reversible measure for this process. The diagonal elements
evolve according to an  OU process with twice variance.
For any $t\ge0$, the  solution to \eqref{Hsde}, $H_t$, has the same distribution as
\be
    e^{-t/2}\wh H + (1-e^{-t})^{1/2} V,
\label{DBMOU}
\ee
where $V$ is a GUE, GOE or GSE matrix.

The generator of the induced stochastic process
on the eigenvalues is given by \eqref{L}.
The equilibrium measure $\mu$ is the GUE, GOE or GSE eigenvalue distribution.
Theorem \ref{thm:main} thus says in this case
that the local eigenvalue statistics
of a Wigner random matrix with a small Gaussian component
coincides with the local statistics of the corresponding Gaussian ensemble.
The entropy condition on $S_\mu(f_{t_0})$ in Theorem \ref{thm:main} can be easily obtained by
\be
   S_\mu(f_{t_0}) \le N^2 S_\gamma(u_{t_0}) \le CN^m.
\label{inentropycond}
\ee
In the real or complex sample covariance case ($\beta=1,2$),
the matrix elements of $A$ evolve according to the OU process \eqref{ou}, i.e.
$A_t$ has the same distribution as
\be
e^{-t/2}\wh A + (1-e^{-t})^{1/2} W,
\ee
where $W$ is an $M\times N$ matrix whose elements are i.i.d
real or complex Gaussian variables with mean 0 and
variance $1/\beta$.

\section{Reverse heat flow}\label{sec:rev}

To remove the short time  restriction from Theorem \ref{thm:main}
in case of Wigner and sample covariance ensembles
and to prove Theorems \ref{thm:main1} and \ref{mainthm2},
we apply
the reverse heat flow argument, presented first in  \cite{ERSY2}
and used also in Corollary 2.4 of \cite{ESY4}.

For fixed $\beta=1,2$ or 4, recall the  Ornstein-Uhlenbeck process from \eqref{ou}
with the reversible  Gaussian measure $\gamma(\rd x)$.
Let $u$ be a positive density with respect to $\gamma$,
i.e., $\int u \rd \gamma=1$
and we write $u(x)=\exp (-V(x))$.
Suppose that for any $K$ fixed  there are constants $C_1, C_2$ depending on $K$ such that
\be
\sum_{j=1}^{2K}  |V^{(j)}(x)| \le C_1  (1+x^2)^{C_2}
\label{cond1}
\ee
and the measure  $\rd \nu = u \rd\gamma $ satisfies the subexponential decay condition.
We will apply this for the initial distribution $\rd\nu = u_0(x) \rd x$,
so $u$ and $u_0$ differ by
a Gaussian factor.

\begin{proposition}\label{meascomp1}
Suppose that $\nu = u \gamma $ satisfies the subexponential decay condition
and \eqref{cond1} for  some $K$. 
 Then there is a small constant $\alpha_K$ depending on $K$ such
that for    $t  \le  \alpha_K $
there exists a probability density $g_t$ with mean zero and variance $\frac{1}{2}$ such that
\be\label{gtilde}
\int  \left  | e^{t B}g_t  - u \right |
\rd \gamma  \le C\; t^{K}
\ee
for  some $C>0$ depending on $K$.  Furthermore,  $g_t$ can be chosen such that
if  the logarithmic Sobolev inequality \eqref{LSI} holds for the measure $\nu=u\gamma$, then
it holds for  $g_t\gamma$  as well, with   the logarithmic Sobolev constant
changing by a factor of at most $2$.

Furthermore, let  $\cB=B^{\otimes n}$, $ F=  u^{\otimes n}$ with some $n\le CN^2$.
Denote by $G_t=  g_t^{\otimes n}$. Then we also have
\be\label{FFtilde}
\int  \left  | e^{t\cB}G_t  - F \right |
\rd \gamma^{\otimes n}   \le C\; N^2 t^{K}
\ee
for  some $C>0$ depending on $K$.
\end{proposition}

We now explain how to prove Theorems \ref{thm:main1} and \ref{mainthm2} 
 from  Theorem \ref{thm:main}
and Proposition \ref{meascomp1}.  
We  choose $n$ to be the number of independent
OU processes needed to generate the flow of the matrix elements.
By choosing $K$ large enough, we can compare  the two measures $e^{t\cB}G_t$ and $ F$
in the total
variational norm; for any observable $J:\bR^n\to \bR$ of the matrix
elements, we have
\[
\left| \int J ( e^{t\cB}G_t  - F )
\rd \gamma^{\otimes n} \right| \le   \|J\|_\infty C\; N^2 t^{K}.
\]
In order to prove Theorems \ref{thm:main1} and \ref{mainthm2}, appropriate observables
$J$ need to be chosen that depend on the matrix elements via the eigenvalues
to express the quantities in \eqref{matrixthm} and \eqref{maineq2}.
It is easy to see that $\|J\|_\infty$ may grow at most polynomially in $N$.
But we can always choose $K$ large enough to compensate for it
with the choice $t= N^{-2\e+\delta}$ allowed in  Theorem \ref{thm:main}.
Here the verifications of the Assumptions I-IV of Theorem \ref{thm:main} were  
explained at the beginning of Section \ref{sec:matr}. 
This completes the
proof of our main theorems. \qed

\medskip
{\it Proof of Proposition \ref{meascomp1}. }   Define $
\theta(  x) =   \theta_0 (  t^{\alpha} x)$ with some  small positive $\alpha>0$
depending on $K$,
where $\theta_0$ is a smooth cutoff function satisfying
$\theta_0(x) = 1$ for $|x|\le 1$ and $\theta_0(x) = 0$ for $|x| \ge 2$.
Set
\[
h_s =    u  +  \theta \xi_s   , \quad \mbox{with} \quad
\; \xi_s:=
\left [ -sB+\frac{1}{2}s^2B^2  + \ldots +  (-1)^{K-1} \frac
{s^{K-1}} { (K-1)!}  B^{K-1}  \right ] u .
\]
By assumption \eqref{cond1},  $h_s$ is positive
and
\be\label{ul}
\frac{2}{3} u  \le h_s \le \frac{3}{2} u.
\ee
for any $s\le t$ if $t$ is small enough.  To see this, take, e.g., $K=2$ 
and we have 
\[
|\theta(  x) \xi_s(x)|  \le C  s  \theta_0 (  t^{\alpha} x) \Big [\,   \big |V''(x) \big | +  \big |x V'(x) \big 
| \, \Big ]   \, | u(x)| \le  \frac{1}{2} | u(x)|,
\]
where we have used $\alpha \ll 1$, $ s \le t$ and the assumption \eqref{cond1}.

Define $  v_s = e^{s B}h_s$ and by definition, $v_0= u$.  Then
$$
\pt_s v_s =   (-1)^{K-1} \frac   {s^{K-1}} { (K-1)!}   e^{s B} B^{K} u
+ e^{s B} B   (\theta -1)  \xi_s + e^{sB} (\theta-1)\pt_s\xi_s.
$$
Since the Ornstein-Uhlenbeck is a contraction in $L^1(\rd\gamma)$,
together with \eqref{cond1}, we have
\be
\int  |v_t - u|  \rd \gamma  \le C_K   \int_0^t     \int  \Big[ t^{K-1}| B^{K} u |  +
|  B   (\theta -1)  \xi_s | +  |(\theta-1)\pt_s\xi_s|  \Big] \rd \gamma \; \rd s
\le  C_K t^{K}
\label{uut}
\ee
for sufficiently small $t$.
To estimate the last two terms, we also used that on the support of
$\theta-1$ the measure $\rd\gamma$ decays subexponentially in $t$.

Notice that $h_t$ may not be normalized as a probability density w.r.t. $\gamma$
but this can be easily adjusted.  To compute this normalization,
 take for example, $K=1$ and we have,  by using $s\le t^\al$,
\[
\Big | \int \theta(  x) \xi_s(x)  \rd \gamma  \Big |  = 
 \Big |   s  \int   \theta_0 (  t^{\alpha} x)   B u(x) \rd \gamma  \Big | 
\le  \Big |     \int     \theta'_0 (  t^{\alpha} x)    u'(x) \rd \gamma  \Big | 
\le     \int_{ |x| \ge   t^{-\alpha/2} }    \Big |     u'(x)   \Big | \rd \gamma.
\]
The last term is bounded by $ O(t^M)$ for any $M > 0$ due to that $u(x)  \gamma$  has a subexponential 
decay and using the assumption  \eqref{cond1} on $V$.

We have proved  that there is a constant $c_t = 1 + O(t^M)$, for any $M > 0$ positive,
such that $c_t h_t$ is a probability density. Clearly,
\[
\alpha_t:= \int x c_t h_t \rd \gamma = O(t^M),   \qquad
\sigma^2_t:=   \int (x-\alpha_t)^2 c_t h_t \rd \gamma = \beta^{-1}+ O(t^M),
\]
and the same formulas hold if $h_t$ is replaced by $v_t$ since
the OU flow preserves expectation and variance.
Let $g_t$ be defined by
\[
g_t(x)  e^{-\beta x^2/2} =  c_t  \sigma^{-1}_t   h_t( (x+\alpha_t) \sigma^{-1} )
e^{-\beta (x+ \alpha_t)^2/2 \sigma^{2}}  .
\]
Then $g_t$ is a probability density w.r.t. $\gamma$ with zero mean and variance $\beta^{-1}$.
It is easy to check that the total variation norm of $h_t-g_t$ is
smaller than any power of $t$. Using again the contraction property of $e^{tB}$ and
\eqref{uut}, we get
\be
\int  |e^{t B} g_t - u|  \rd \gamma
\le  C t^{K}
\label{uut1}
\ee
for sufficiently small $t$.

Now we  check the LSI constant for $g_t$.  Recall that $g_t$ was
obtained from $h_t$ by translation and dilation. By definition of the LSI
constant, the translation does not change  it.
The dilation changes the constant, but since our dilation constant is nearly one,
the change of LSI constant is also nearly one.
So we only have to compare the LSI constants between $\rd\nu = u\rd\gamma$ and $
c_t   h_t \rd \gamma  $. From \eqref{ul} and that $c_t$ is nearly one,
the LSI constant changes by a factor less than $2$.
This proves the claim on the LSI constant.

Finally, the \eqref{FFtilde} directly follows from
$$
\int  \left  | e^{t\cB}G_t  - F \right |
\rd \gamma^{\otimes n} \le n \int  \left  | e^{tB}g_t  - u \right |\rd\gamma
$$
and this completes the proof of Proposition \ref{meascomp1}.
\qed

\section{Proof of Theorem \ref{thm:main}}\label{sec:corfn}

We start with the identity
\begin{align}
\int_{E-b}^{E+b}\rd E' \int_{\R^n} & \rd\alpha_1
\ldots \rd\alpha_n   \; O(\alpha_1,\ldots,\alpha_n)  p_{\tau, N}^{(n)}
\Big (E'+\frac{\alpha_1}{N\varrho(E)},
\ldots, E'+\frac{\alpha_n}{N \varrho(E)}\Big) \label{iddd} \\
 = & C_{N,n}\int_{E-b}^{E+b}\rd E' \int \sum_{i_1\ne i_2\ne
 \ldots \ne i_n}
  \wt O\big( N(x_{i_1}-E'),  N(x_{i_1}-x_{i_2}),
 \ldots  N(x_{i_{n-1}}-x_{i_n})\big)
 f_\tau\rd\mu, \non
\end{align}
where $\wt O (u_1, u_2, \ldots u_n): = 
O\big( \varrho(E)u_1, \varrho(E)(u_2-u_1), \ldots\big)$
 and $C_{N,n} = N^n (N-n)!/N!= 1 + O_n(N^{-1})$.
By permutational symmetry of
$ p_{\tau, N}^{(n)}$ we can assume that $O$ is symmetric and we can restrict
the last summation to $i_1 < i_2 < \ldots < i_n$ upon an overall factor $n!$.
Let $S_n$ denote the set of  
increasing positive integers, ${\bf m} = (m_2, m_3, \ldots, m_n) \in \N_+^{n-1}$, 
$m_2< m_3 <\ldots < m_n$.
For a given $\bm\in S_n$, we
change indices to $i= i_1$, $i_2= i+m_2$, $i_3=i+m_3, \ldots,$
 and rewrite
the sum on the r.h.s. of \eqref{iddd} as
\begin{align}
  \sum_{{\bf m}\in S_n} \sum_{i=1}^N & \wt O\big(  N(x_i - E'),
N(x_i - x_{i+m_2}),
  N(x_{i+m_2} - x_{i+m_3}) , \ldots \big) =  \sum_{{\bf m}\in S_n}\sum_{i=1}^N  Y_{i,\bm}
(E',\bx),
\non 
\end{align}
where we introduced
$$
   Y_{i,\bm}(E',\bx) = \wt O\big(  N(x_i - E'),
N(x_i - x_{i+m_2}) ,\ldots,  N(x_i- x_{i+m_n})   \big).
$$
We will set $Y_{i,\bm}=0$ if $i+m_n>N$.
 Our goal is to estimate the difference 
\be
\Theta:=  
  \Bigg| \int_{E-b}^{E+b}\frac{\rd E'}{2b} \int \sum_{{\bf m}\in S_n}\sum_{i=1}^N  Y_{i,\bm}
(E',\bx) (f_\tau-1) \rd \mu \Bigg|.
\label{goal}
\ee
Let $M$ be an $N$-dependent parameter chosen at the end of the proof,
in fact, $M$ will be chosen as
$M=N^c$ with some small positive exponent $c>0$, depending on $n$.
 Let 
$$
  S_n(M): = \{ \bm \in S_n \; , \; m_n \le M\}, \quad S_n^c(M):= S_n\setminus S_n(M),
$$
and note that $|S_n(M)|\le M^{n-1}$.  We have the simple bound $\Theta \le  \Theta_M^{(1)}(\tau) + 
\Theta_M^{(2)}(\tau) + \Theta_M^{(2)}(\infty) $ where 
\be
 \Theta_M^{(1)}(\tau):  =   
  \Bigg| \int_{E-b}^{E+b}\frac{\rd E'}{2b} \int \sum_{{\bf m}\in S_n(M)}  \sum_{i=1}^N  Y_{i,\bm}
(E',\bx) (f_\tau-1)\rd \mu   \Bigg| 
\label{goal3}
\ee
and
\be
 \Theta_M^{(2)}(\tau):  =   \sum_{{\bf m}\in S_n^c(M)}
  \Bigg| \int_{E-b}^{E+b}\frac{\rd E'}{2b} \int \sum_{i=1}^N  Y_{i,\bm}
(E',\bx) f_\tau\rd \mu   \Bigg| 
\label{goal2}
\ee
Note that $\Theta_M^{(2)}(\infty)$ is the same as $\Theta_M^{(2)}(\tau)$ but with $f_\tau$ replaced by 
the constant $1$, i.e., $f_\infty \rd \mu$ is the equilibrium.

\bigskip

\noindent
{\it Step 1: Small $\bm$ case}

\bigskip
 After performing the $\rd E'$ integration, we will eventually
apply Theorem \ref{thmM} to the function
$$
G\big( u_1, u_2, \ldots \big)
 : = \int_{\R} \wt O\big( y,
 u_1, u_2 ,\ldots,   \big) \rd y ,
$$
i.e., to the quantity
\be
 \int_\R \rd E'\; Y_{i,\bm}(E',\bx)= 
 \frac{1}{N} G\Big( N(x_i-x_{i+m_2}), \ldots \Big) 
\label{OO}
\ee
for each fixed $i$ and ${\bf m}$.

For any $E$ and  $0<\xi<b$ define sets of integers
$J=J_{E,b,\xi}$ and $J^\pm= J^\pm_{E,b,\xi}$  by
$$
  J : = \big\{ i\; : \; \gamma_i \in [E-b, E+b]\big\},\quad
  J^\pm : = \big\{ i\; : \; \gamma_i \in [E-(b\pm\xi), E+b\pm\xi]\big\}, 
$$
where $\gamma_i$ was defined in \eqref{gammaj}.  Clearly $J^-\subset J \subset J^+$.
With these notations, we have
\be
   \int_{E-b}^{E+b} \frac {\rd E'} {2b} \sum_{i=1}^N
 Y_{i,\bm}(E',\bx )=   \int_{E-b}^{E+b} \frac {\rd E'} {2b}  \sum_{i\in J^+} Y_{i,\bm}(E',\bx )
 + \Omega^+_{J,\bm}(\bx).
\label{uppe}
\ee
 The error term $\Omega^+_{J,\bm}$, defined by \eqref{uppe}
indirectly, comes from  those $i\not\in J^+$ indices,
for which $x_i \in [E-b+ O(N^{-1}), E+b+ O(N^{-1})] $
since 
$Y_{i,\bm}(E',\bx)=0$ unless $|x_i-E'|\le C/N$, the constant
depending on the support of $O$. Thus
\be\label{1.133}
   |\Omega^+_{J,\bm}(\bx)| \le  C  b^{-1} N^{-1}\# \{ \; i \; : \; |x_i-\gamma_i|\ge \xi/2 \}
\ee
for any sufficiently large $N$,  assuming $\xi\gg 1/N$
and using that $O$ is a bounded function. The additional $N^{-1}$ factor
comes from the $\rd E'$ integration. 
 Taking the expectation with respect to the
measure $f_\tau\rd\mu$, we get
\be
    \int |\Omega^+_{J,\bm}(\bx)| f_\tau\rd\mu \le C  b^{-1} \xi^{-2}N^{-1} 
 \int \sum_i (x_i-\gamma_i)^2 f_\tau\rd\mu 
   = C  b^{-1} \xi^{-2} N^{-1-2\e}
\label{exp}
\ee
using Assumption III \eqref{assum2}.
We can also estimate
\begin{align}
&   \int_{E-b}^{E+b}\frac {\rd E'} {2b} \sum_{i\in J^+} Y_{i,\bm}(E',\bx) \nonumber  \\
  \le &  \int_{E-b}^{E+b} \frac {\rd E'} {2b} \sum_{i\in J^-} Y_{i,\bm}(E',\bx) + 
Cb^{-1} N^{-1}  |J^+\setminus J^-|  \non \\
   = &  \int_\R \frac {\rd E'} {2b}  \sum_{i\in J^-}Y_{i,\bm}(E',\bx)  + C  b^{-1} N^{-1}|J^+\setminus J^-|+
  \Xi^+_{J,\bm}(\bx) \label{fol} \\
 \le &  \int_\R \frac {\rd E'} {2b}  \sum_{i\in J}Y_{i,\bm}(E',\bx)+
Cb^{-1} N^{-1}|J^+\setminus J^-|+C b^{-1}N^{-1}|J\setminus J^-|+
  \Xi^+_{J,\bm}(\bx), \non
\end{align}
where the error term $\Xi^+_{J,\bm}$, defined by \eqref{fol}, 
comes from indices $i\in J^-$ such that $x_i \not \in [E-b, E+b]+O(1/N)$.
It satisfies the same bound \eqref{exp} as $\Omega^+_{J,\bm}$.

By the continuity  of $\varrho$, the density of $\gamma_i$'s is
bounded by $CN$, thus $|J^+\setminus J^-|\le CN\xi$ and
$|J\setminus J^-|\le CN\xi$.
 Therefore, 
summing up the formula \eqref{OO} for $i\in J$,
 we obtain from \eqref{uppe} and \eqref{fol}
\begin{align}
   \int_{E-b}^{E+b} \frac {\rd E'} {2b}  \int & \sum_{i=1}^N
Y_{i,\bm}(E',\bx)  f_\tau\rd\mu \\ \le  \, &   (2b)^{-1}   \int \frac{1}{N}\sum_{i\in J}
 G \Big( N(x_i-x_{i+m_2}), \ldots \Big)
    f_\tau\rd\mu + C b^{-1}  \xi + C  b^{-1}  \xi^{-2} N^{-1-2\e}
\end{align}
for each $\bm\in S_n$.
A similar lower bound can be proved analogously and we obtain
\begin{align}
  \Bigg|     \int_{E-b}^{E+b} \frac {\rd E'} {2b} \int \sum_{i=1}^N
 Y_{i,\bm}(E',\bx) f_\tau\rd\mu   \nonumber  -    (2b)^{-1}   \int \frac{1}{N}
\sum_{i\in J} G \Big( & N(x_i-x_{i+m_2}), \ldots \Big)
    f_\tau\rd\mu \Bigg| \\
&\, \le  C b^{-1}  \xi + C  b^{-1}  \xi^{-2} N^{-1-2\e}
    \label{seccc}
\end{align}
for each $\bm\in S_n$.

 Adding up \eqref{seccc}
for all $\bm\in S_n(M)$, we get
\begin{align}
   \Bigg|  &  \int_{E-b}^{E+b}  \frac {\rd E'} {2b}   \int \sum_{\bm\in S_n(M)}\sum_{i=1}^N
 Y_{i,\bm}(E',\bx) f_\tau\rd\mu  \nonumber  \\ &  -    (2b)^{-1}   \int \sum_{\bm\in S_n(M)} \frac{1}{N}
\sum_{i\in J} G  \Big( N(x_i-x_{i+m_2}), \ldots \Big)
    f_\tau\rd\mu \Bigg|    
    \, \le  C b^{-1}  \xi + C  b^{-1}  \xi^{-2} N^{-1-2\e} ,
\label{seccc1}
\end{align}
and the same estimate holds for the equilibrium, i.e.,
if we set $\tau=\infty$ in \eqref{seccc1}. 
We now subtract the  these  two formulas and apply \eqref{GG} from
 Theorem \ref{thmM}
to each summand on the second term in \eqref{seccc1}. Choosing  
$\xi= N^{-(1+ 2 \e) /3}$ to minimize the two error terms involving $\xi$,  we conclude 
that
\begin{align}
\Theta^{(1)}_M   =  &     \Bigg|  \int_{E-b}^{E+b} \frac {\rd E'} {2b} 
 \int \sum_{\bm\in S_n(M)}\sum_{i=1}^N
 Y_{i,\bm}(E',\bx) (f_\tau\rd\mu -\rd \mu) \Bigg|  \nonumber 
 \\   & \le  CM^{n-1}\Big( b^{-1}  N^{ - \frac { 1 +  2 \e} 3 } +  b^{-1/2} N^{\e'-\delta/2} \Big).
\label{ssd}
\end{align}
where we have used  $\tau = N^{-2 \e + \delta}$  and that $|J| \le CNb$.

\bigskip
\noindent
{\it Step 2. Large $\bm$ case.}
\bigskip

For a fixed $y\in \R$, $\ell >0$, let
$$
   \chi(y,\ell) : =\sum_{i=1}^N {\bf 1}\Big\{ x_i\in \big[ y- \frac{\ell}{N},
 y +\frac{\ell}{N}\big] \Big\}
$$
denote the number of points in the interval $[y-\ell/N, y+\ell/N]$.
 Note that for a fixed $\bm=(m_2, \ldots , m_n)$, we have
\be
   \sum_{i=1}^N |Y_{i,\bm} (E',\bx)| \le C\cdot\chi(E',\ell)
 \cdot {\bf 1}\Big(\chi(E',\ell)\ge m_n\Big) \le C\sum_{m=m_n}^\infty m \cdot
 {\bf 1}\Big(\chi(E',\ell)\ge m\Big),
\label{Ofi}
\ee
where $\ell$ denotes the maximum of $|u_1|+\ldots + |u_n|$
in the support of  $\wt O(u_1, \ldots , u_n)$.

 Since the summation over
all increasing sequences
$\bm = (m_2, \ldots, m_n)\in \N_+^{n-1}$ with a fixed $m_n$
contains at most $m_n^{n-2}$ terms,
by definition \eqref{goal2} we have 
\be
 \Theta_M^{(2)}(\tau)    \le C \int_{E-b}^{E+b} \frac {\rd E'} {2b}  \;\sum_{m=M}^\infty m^{n-1}
  \int  {\bf 1}\Big(\chi(E',\ell)\ge m\Big) f_\tau\rd\mu.
\label{toc}
\ee
Now we use  Assumption IV for the interval
$I = [E' - N^{-1+\sigma}, E' + N^{-1+\sigma}]$ with 
$\sigma$ chosen in such a way that $N^{\sigma} \le M^2$.
Clearly $\cN_I\ge \chi(E',\ell)$ for
sufficiently large $N$, thus we get from  \eqref{ass4} that
$$
   \sum_{m=M}^\infty m^{n-1}
  \int  {\bf 1}\Big(\chi(E',\ell)\ge m\Big) f_\tau\rd\mu 
 \le C_a \sum_{m=M}^\infty m^{n-1} \Big(\frac{m}{N^\sigma}\Big)^{-a} 
$$
holds for any $a\in \N$. By the choice of $\sigma$,
we get that $\sqrt{m}\ge N^\sigma$ for any $m\ge M$,
and thus choosing $a=k(n+1)$, we get
$$
 \Theta_M^{(2)}(\tau)    \le \frac{C_a}{M^{k-1}}. 
$$
Together with \eqref{ssd}, we have thus proved that 
\be
\Theta \le  CM^{n-1}\Big( b^{-1}  N^{ - \frac { 1 +  2 \e} 3 } +
  b^{-1/2} N^{\e'-\delta/2} \Big)+ \frac{C_a}{M^{k-1}}.
\ee
Choosing $M$ such that $M^n = N^{\e'}$ and then choose $k$ large enough so that the last term 
$\frac{C_a}{M^{k-1}} $ is smaller than, say, $N^{-2}$. We have thus proved that 
\be\label{7.7}
\Theta \le  C N^{2 \e'} [ b^{-1}  N^{ - \frac { 1 +  2 \e} 3 } +  b^{-1/2} N^{-\delta/2} ]
\ee
for $\tau = N^{-2 \e + \delta}$ and this  
concludes   \eqref{abstrthm}.

For the proof of  \eqref{abstrthm2}, we choose $\xi \ge 2N^{-1 + A}$,
and then by using   \eqref{memory}    we can estimate 
$\Omega^+_{J,\bm}$ directly as
\be
    \int |\Omega^+_{J,\bm}(\bx)| f_\tau\rd\mu \ll N^{-K}
\label{exp1}
\ee
for any $K>0$, instead of \eqref{exp}.
Therefore, the estimate on the right hand side 
of \eqref{seccc} and the subsequent estimates  can be replaced by 
\be
C b^{-1}  \xi + C  b^{-1} N^{-K} 
\ee
provided $\xi \ge 2N^{-1 + A}$. Choosing  $\xi = 2N^{-1 + A}$ 
and following the same  proof, we can improve the estimate \eqref{7.7}  to
\be\label{7.8}
\Theta \le  C N^{2 \e'} [ b^{-1} N^{-1 + A}+  b^{-1/2} N^{-\delta/2} ]
\ee
for $\tau = N^{-2 \e + \delta}$. This proves  \eqref{abstrthm2} and we have completed 
the proof of Theorem \ref{thm:main}.

\qed

\section{Local Marchenko-Pastur law}\label{sec:wish}

In this section we establish that
the empirical density of eigenvalues for sample covariance
matrices is close to the Marchenko-Pastur law even
on short scale. We do this by controlling the
difference of the Stieltjes transform, establishing
results analogous to Theorem 4.1. and Proposition 4.2
of \cite{ERSY}.
In this section, we focus on $0<d=N/M<1$, in particular the lower
spectral edge $\lambda_->0$. The constants appearing in this subsection may depend on $d$.

Before the detailed proof, we explain the main steps of the argument
which  is  similar to the method we have successively developed in \cite{ESY1, ESY2, ESY3}. 
The proof given here  is somewhat complicated by fact that the matrix elements themselves are not
independent but are generated as a quadratic expression of independent random variables.
 The first step, Lemma \ref{lm:up}, is an apriori bound on the 
local density on short scales, $\eta \gg 1/N$, using resolvent expansion
and a large deviation principle for quadratic forms.
Expressing the resolvent of $H$ in terms the resolvents
of its minors, we obtain a self-consistent equation \eqref{IDMzs}
for the Stieltjes transform $m_N$ of the eigenvalues. This equation
is very close to the defining quadratic equation 
of the Stieltjes transform $m_W$ of the Marchenko-Pastur law, see \eqref{defmW},
with a perturbation term $Y(z)$.
This term can be estimated by large deviation arguments and using 
the a-priori bound on the local density. 
Then in Lemma \ref{mNmWonline} we investigate the stability of
the self-consistent equation for $m_W$. Although the perturbed equation
has two solutions, only one of them can be close to $m_N$. To select
the correct solution, we use a continuity argument in the spectral
parameter $z$. For $z=z_0$ with a large imaginary part, say $z_0= 10+5i$,
the explicit formula \eqref{EExpl} for the solution can be directly 
analyzed. For $z$ approaching to the real axis, we prove that 
the two unperturbed solutions remain  far away from each other \eqref{temp6.21}.
Since the perturbed solutions are also continuous in the spectral parameter,
for a sufficiently small perturbation they must remain in the vicinity
of the correct solution of the unperturbed equation.

This analysis yields a bound on the difference of Stieltjes transforms,
$m_N - m_W$. In Lemma \ref{EMg2} we give a better bound on 
$\E m_N -m_W$. The improvement is due to the fact that 
the perturbation term $Y$ in the self-consistent 
equation is random and its expectation is much smaller than its
typical size (compare \eqref{PYJ} and \eqref{temp7.31}).
Finally, in Lemma \ref{EMg1} we give an independent estimate
on $\E m_N -m_W$ that is weaker  in terms of $\eta=\text{Im}z$ but
it is weaker in $\kappa$. When we will
verify  Assumption III in the following Section \ref{sec:check},
we will use both bounds simultaneously.

\begin{lemma}\label{lm:up}
Let $0< E<10$ and $0< d<1$. Consider the interval $I_{\eta} = [ E -\eta , E + \eta]$.
Let $\cN_I$ denote the number of eigenvalues of $H = A^* A$ in the interval $I_\eta$.
Suppose that $ N^{-1+ \eps} \leq \eta \leq E/2$, for some $\eps >0$. Then
there exist constants $C,c>0$ such that
\be\label{upp} \P \left( \cN_{I_\eta} \geq \frac{K N \eta }{\sqrt{E}} \right)
\leq C e^{-c \sqrt{KN\eta/\sqrt E} },\ee
for all $N, K$ large enough (independent of $E$).
\end{lemma}
\par We remark that the assumption on $\eta$ can be relaxed  to $ CN^{-1} \leq \eta \leq E/2$.
But we do not need this result here. For details, one can refer to Theorem 5.1 of \cite{ESY3}.

\medskip
\par {\it Proof of Lemma \ref{lm:up}. }
We observe, first of all, that
\[ \frac{\cN_I}{N\eta} \leq \frac{C}{N} \text{Im} \, \tr \; \frac{1}{H-E -i \eta} = \frac{C}{N}
\text{Im} \, \sum_{j=1}^N \frac{1}{H-z} (j,j), \] where we defined $z= E +i\eta$.
It follows that \begin{equation}\label{eq:up1} \P \left( \cN_I \geq KN\eta /\sqrt{E}\right)
\leq N \P \left( \left| \text{Im} \, \frac{1}{H-z} (1,1) \right| \geq K/\sqrt{E} \right).\end{equation}

Denoting by $a_1$ the first column of $A$ and by $B$ the $M \times (N-1)$ matrix
consisting of the last $N-1$ column of $A$, we have
\[ H = \left( \begin{array}{ll} a_1 \cdot a_1 & (B^* a_1 )^* \\ B^* a_1 & B^* B \end{array} \right). \]
Hence \[ \frac{1}{H-z} (1,1) = \frac{1}{a_1\cdot a_1 -  z - a_1 \cdot B (B^* B -z)^{-1} B^* a_1}. \]
Using the identity \[ B (B^* B -z)^{-1} B^* = B B^* (B B^* -z)^{-1}, \]
we find
\be\label{Hzaabb*}
\frac{1}{H-z} (1,1) = \frac{1}{a_1\cdot a_1 -  z - a_1 \cdot B B^* (B B^* -z)^{-1} a_1}.
\ee
Denote $\mu_\alpha$'s $(\alpha=1,\ldots, N-1)$ the eigenvalues of the
$(N-1) \times (N-1)$ matrix $B^* B$. The $\mu_\alpha$'s are also the eigenvalues of
$M \times M$ matrix $B B^*$ and the other eigenvalues of  $B B^*$ are zeros.
Then define $v_{\alpha}$ $(\alpha=1,\ldots, N-1)$ as  the normalized eigenvectors
of $BB^*$ associated with non-zero eigenvalues $\mu_\alpha$, i.e.,
 the matrix elements of $BB^*$ are given by
\be\label{BBluu}
(BB^*)_{ij}=\sum_{\alpha=1}^{N-1}\mu_\alpha \bar v_\alpha(i)v_\alpha(j).
\ee
Inserting \eqref{BBluu} into \eqref{Hzaabb*},  we find
\[ \frac{1}{H-z} (1,1) = \frac{1}{a_1\cdot a_1 -  z - \frac{1}{M} \sum_{\alpha=1}^{N-1}
\frac{\mu_\alpha \xi_\alpha}{\mu_\alpha -z}}\,,\]
where we defined the quantity $\xi_\alpha = M |a_1 \cdot v_\alpha|^2$ (note that
$\E \xi_\alpha = 1$). Taking the imaginary part, we find
\[ \begin{split}
\left| \, \text{Im} \, \frac{1}{H-z} (1,1) \right| \leq \frac{1}{\eta + \frac{\eta}{M}
\sum_{\alpha=1}^{N-1} \frac{\mu_\alpha \xi_\alpha}{(\mu_\alpha-E)^2 + \eta^2}}
\leq \frac{C N\eta}{E \, \sum_{\alpha: |\mu_\alpha - E| \leq \eta} \xi_\alpha} \,\,,
\end{split} \]
where we used the assumption that $\eta < E/2$. Because the eigenvalues
$\mu_\alpha$'s are  the eigenvalues of the
$(N-1) \times (N-1)$ matrix $B^* B$, they are interlaced with the
eigenvalues of $H$ and $| \{ \alpha : |\mu_\alpha - E| \leq \eta/2 \} |
\geq \cN_I -1$. It follows from (\ref{eq:up1}) that
\[ \P \left( \cN_I \geq \frac{KN\eta}{\sqrt{E }} \right)
\leq N \P \left( \sum_{\alpha: |\mu_\alpha -E|\leq \eta/2}
\xi_\alpha  \leq \frac{CN\eta}{ K\sqrt{E}} \text{ and }
\cN_I \geq \frac{KN\eta}{\sqrt{E}}  \right) \leq C N e^{-c\sqrt{KN\eta/\sqrt E}}\,\,, \]
where in the last step we used Lemma 4.7 from \cite{ESY3}.
The claim follows by the assumption that $N\eta \geq N^{\e}$ and that $N$ and $K$ are large enough.
\qed
\begin{proposition} \label{mainprop}
Consider sample covariance matrices $H=A^*A$ with $A$ an $M \times N$ matrix with independent and
identically distributed complex entries. Let $0< d<1$. Recall $\lambda_-$ and $\lambda_+$
in \eqref{deflambda-+} and define $\kappa$ as
\begin{equation}\label{def:kappa}
\kappa=\kappa(E):=\left|(E-\lambda_-)(E-\lambda_+)\right|.
\end{equation}
We will often drop the argument $E$ from the notation
of $\kappa$ for brevity.
Then for any $E$, $\eta$ satisfying $N^{-1+\eps}\leq \eta\leq \frac12E$, 
$\frac12\lambda_-\leq E\leq 10$, the Stieltjes transform,
\[ \mg (z): = \frac{1}{N} \tr\, \frac{1}{H-z}\,,\,\,\, z=E+i\eta,\] of the empirical
eigenvalue distribution of  $H = A^* A$ satisfies
\be\label{PmNmW}
\P \left( \left|\mg(E+i\eta) - \mW (E +i\eta) \right|
 \geq {\frac{\delta }{\sqrt{\kappa+\delta}}} \right)
\leq C e^{-c \delta \sqrt{N\eta}}\,\,,
\ee for any $\delta$ small enough (independent of $E$ and $\eta$) and $N \geq 2$.
Here $m_W (z)$ is the unique solution of
\be\label{defmW} \mW (z)  + \frac{1}{z  - (1-d) + z\, d \,\mW (z)} = 0,
\ee
with positive imaginary part for all $z$ with $\text{Im } z > 0$.
\end{proposition}
Recall $\lambda_\pm=(1\pm d^{1/2})^2$ from \eqref{deflambda-+}.  The function
$m_W$ defined in \eqref{defmW}  depends on $d$ and can be written as
\begin{equation}\label{mwz=}
m_W(z)=\frac{1-d-z +i \sqrt{(z-\lambda_-)(\lambda_+-z)}}{2\,d\,z},
\end{equation}
where $\sqrt{\,\,}$ denotes the square root on complex 
plane whose branch cut is the negative real line.
Explicit calculation shows $m_W(z)$ is  the  Stieltjes transform of the Marchenko-Pastur
density
given in \eqref{defrhow}.

Using \eqref{PmNmW} and \eqref{defrhow}, we have the local Marchenko-Pastur law
for the number of eigenvalues in a small interval:
\begin{corollary}\label{localMPlaw}
Consider an interval $I=[E-\eta, E+\eta]\subset [\lambda_-,\lambda_+]$ within
the bulk spectrum. Let $\delta$ be a suffiiciently small parameter.
Suppose that
$E$ and $\eta$ are chosen such that   $\delta^{-2} N^{-1+\e}\leq \eta\leq C^{-1}
\min\{ \kappa, \delta^{1/2}\kappa^{3/4}\}$ with a large constant $C$ and
with $\kappa= \kappa(E)$ given in  \eqref{def:kappa}.
Then we have the convergence
of the counting function, i.e.,
\be\label{reslocalMPlaw}
\P\left\{\left|\frac{\mathcal N_{\eta}(E)}{2\eta N}-\rho_W(E)\right|\geq \delta\right\}\leq
C e^{-c\delta^2\sqrt{N\eta\kappa} },
\ee
where $\mathcal N_{\eta}(E) = |\{\lambda_\alpha: |\lambda_\alpha-E|\leq \eta\}|$
denotes the number of eigenvalues of $H=A^*A$ in the interval $I=[E-\eta,E+\eta]$.
\end{corollary}

\par {\it Proof of Corollary \ref{localMPlaw}.} The proof of \eqref{reslocalMPlaw} follows
from  the inequality
\eqref{PmNmW}  with a similar argument as the proof of Proposition 4.1 of \cite{ERSY}. \qed

\medskip

We remark that, similarly to  Theorem 3.1 in \cite{ESY3},
the assumption on the lower bound $\eta\ge N^{-1+\e}$ can be relaxed  to $\eta\ge  KN^{-1}$
and obtain the local Marchenko-Pastur law on the shortest possible scale, at least away
from the spectral edges.

\bigskip
\par {\it Proof of Proposition \ref{mainprop}.}
Let $a_j$ be the $j$-th column of $A$ and let  $B^{(j)}$ be  the remaining
$M\times (N-1)$ matrix obtained from $A$ after removing the $j$-th column $a_j$.
Let  $\mu^{(j)}_\alpha$, $v^{(j)}_\alpha$ be the non-zero eigenvalues and the eigenvectors
of  the matrix $B^{(j)}[B^{(j)}]^*$ and we define  $\xi_\alpha^{(j)}= M |a_j \cdot v^{(j)}_\alpha|^2$.
Then we have the formula
\[ \mg(z) = \frac{1}{N} \tr \, \frac{1}{H-z} = \frac{1}{N} \sum_{j=1}^N \frac{1}{H-z} (j,j)
= \frac{1}{N} \sum_{j=1}^{N} \frac{1}{a_j \cdot a_j - z -
\frac{1}{M} \sum_{\alpha=1}^{N-1} \frac{\mu^{(j)}_\alpha}{\mu^{(j)}_\alpha -z} \xi^{(j)}_\alpha} \]
that we rewrite as
\[ \begin{split}
\mg (z) &=
\frac{1}{N} \sum_{j=1}^N \frac{1}{a_j \cdot a_j - z - \frac{N-1}{M}- \frac{z}{M}
\sum_{\alpha=1}^{N-1} \frac{1}{\mu^{(j)}_\alpha -z} - X^{(j)}}, \end{split} \]
with \be\label{defXj} X^{(j)}=X^{(j)}(z) =  \frac{1}{M} \sum_{\alpha=1}^{N-1}
\frac{\mu^{(j)}_\alpha}{\mu^{(j)}_\alpha -z} \left(\xi^{(j)}_\alpha - 1 \right). \ee
Note that the vector $a_j$ is independent of $\mu^{(j)}_\alpha$ and $v^{(j)}_\alpha$.
 Therefore, we have
\be\label{EXj0}
\E X^{(j)}=0,\,\,\, \E\xi^{(j)}_\alpha=1.
\ee
Define $ m_{N-1}^{(j)} (z)\equiv \frac{1}{N-1}{\rm Tr} ([B^{(j)}]^*B^{(j)}-z)^{-1}$, then
\[ \sum_{\alpha=1}^{N-1}  \frac{1}{\mu^{(j)}_\alpha -z}
= (N-1) m_{N-1}^{(j)} (z)\,.
\]
Hence
\begin{eqnarray}
\label{IDMz}
\mg (z) =
\frac{1}{N} \sum_{j=1}^N \frac{1}{1-z -d -z \,d\, m_N (z)  + Y^{(j)}}\,\,,\end{eqnarray}
with
\be\label{defYj} Y^{(j)}=Y^{(j)}(z) = (a_j \cdot a_j - 1) + \frac{1}{M} -
\frac{z}{M} \left( (N-1) m_{N-1}^{(j)} (z) - N\mg (z) \right) - X^{(j)}(z).
\ee
For fixed $j$, denote $b = \sqrt{M} a_j$ with $b=(b_1,\ldots,b_M)$. Drop the superscript
$j$ for $\mu_\alpha=\mu_\alpha^{(j)}$, $v_\alpha=v_\alpha^{(j)}$, $B=B^{(j)}$
and $m_{N-1}=m_{N-1}^{(j)}$ for simplicity. We rewrite $X^{(j)}$ as
$$
 X^{(j)} = \sum_{\ell,k=1}^{M} \sigma_{\ell k} [ b_\ell \bar b_k - \E b_\ell \bar b_k]
$$
with
$$
\sigma_{\ell k}: =\frac{1}{M}\sum_{\al=1}^{N-1} \frac{\mu_\al \bar v_\al(\ell) v_\al(k)}{\mu_\al-z}.
$$
So with $\frac12\lambda_-\le E=Re (z)\le 10$ and $N^{-1+\eps}\leq \eta\leq E/2$, we have
$$
\sum_{\ell,k} |\sigma_{\ell k}|^2
= \frac{1}{M^2} \sum_\al \frac{\mu_\al^2}{|\mu_\al-z|^2}
\le\frac{CK}{M\eta} ,
$$
with some fixed large $K$
(using dyadic decomposition and  \eqref{upp},  similarly to the argument in Lemma 4.2 of \cite{ESY3})
apart from an event of probability $e^{-c\sqrt{N\eta}}$.
Then with Proposition 4.5 of \cite{ESY3}, we have

\begin{equation}\label{eq:err1}
\P (\max_j|X^{(j)}|\ge \delta ) \le Ce^{-c \delta \sqrt{N\eta}},
\end{equation}
for sufficiently small $\delta>0$.
Since the eigenvalues of $B^*B$ are interlaced with the eigenvalues of $H=A^*A$, we have
\begin{equation}\label{eq:err2}
\left| (N-1) m_{N-1} (z) - N m_N (z) \right| \leq C \eta^{-1}.
\end{equation}
Then using $\E a_j\cdot a_j=\frac1M\E\sum_{i=1}^M|b_i|^2=1$ and Proposition 4.5 of \cite{ESY3}
for the iid variables $b_i$'s, we obtain
\begin{equation}\label{eq:err3} \P \left( |a_j \cdot a_j - 1|
\geq K M^{-1/2} \right) \leq Ce^{-c\min\{K,K^2\}}. \end{equation}
Combining (\ref{eq:err1}), (\ref{eq:err2}), and (\ref{eq:err3}), we find that
\[ \P \left(\left|Y^{(j)}(z)\right| \geq \delta \right) \leq C e^{-c\delta \sqrt{N\eta}}
+Ce^{-c\delta^2 N} \,,  \]
for sufficiently small $\delta>0$. With the assumption $\eta\leq Re(z)/2=E/2$
and $N\eta\geq N^\e$, we obtain
\be\label{PYJ} \P \left(\max_j\left|Y^{(j)}(z)\right| \geq \delta \right)
\leq C e^{-c\delta \sqrt{N\eta}} \, . \ee
On the other hand, with the definition of $Y^{(j)}$, for  any $j$, $z$, $z'$ such that,
$|z|,|z'|\leq 10$, $\mbox{Im}(z), \mbox{Im}(z')\geq \eta$, we have
\begin{equation}
\P\left(\left|Y^{(j)}(z)-Y^{(j)}(z')\right|\geq \eta^{-2}|z-z'| \right)\leq e^{-c\sqrt{N\eta}}.
\end{equation}
Together with \eqref{PYJ}, we obtain, for $N^{-1+\eps}\leq \eta\leq \frac12E$,
$\frac12\lambda_-\leq E\leq 10$ and sufficiently small $\delta$,
\be\label{PYJ2} \P \left(\max_{z'\in L(z,P_{10})}\max_j\left|Y^{(j)}(z')\right|
\geq \delta \right) \leq C e^{-c\delta \sqrt{N\eta}} \, ,
\,\,\, P_{10}=10+5 i,\ee
where $L(z,P_{10})$ is the line segment connecting points $z$ and $P_{10}$.
Then  the Proposition \ref{mainprop} follows from  the next lemma.
\qed
\bigskip
\begin{lemma}\label{mNmWonline} Assume  $H$ is a $N\times N$  positive semidefinite matrix with
 $\|H\|\leq 5$.
 For fixed $0<d<1$,
we recall the notation $\lambda_\pm=(1\pm \sqrt d)^2$.
Let $z_0=E+i\eta $ and $N^{-1+\eps}\leq \eta\leq \frac12E$, $\frac12\lambda_-\leq E\leq 10$.
Denote $L(z_0,P_{10})$ the line segment connecting $z_0$ and $P_{10}=10+5i$.  Suppose
that for any $z\in L(z_0,P_{10})$, the Stieltjes transform
$m_N(z)=\frac{1}{N}{\rm Tr} (H-z)^{-1}$  satisfies the following self-consistent relation:
\be\label{IDMzs}
\mg (z) =
\frac{1}{N} \sum_{j=1}^N \frac{1}{1-z -d -z \,d\, m_N (z)  + Y^{(j)}(z)},
\ee
for some $Y^{(j)}(z)$'s.
Then there exists $\delta_0>0$  depending only on $d$, such that, whenever
\be\label{temp6.14}
\delta\equiv \max_{z\in L(z_0,P_{10})}\max_j\left|Y^{(j)}(z)\right|  \le \delta_0,
\ee
we have
\be\label{resultmNmWonline}
\left|\mg(z_0)-\mW(z_0)\right|\leq C \delta(\kappa+\delta)^{-1/2}
\ee
with $ \kappa =\kappa(E):=|(\lambda_+-E)(E-\lambda_-)|$.
\end{lemma}

\par {\it Proof of Lemma \ref{mNmWonline}.}
We begin with a special case: $z_0=P_{10}$. In this case if $z\in L(z_0, P_{10})$
then $z=z_0=P_{10}$.
With the assumptions on $H$ and $0<d<1$, it is easy to see that:
 \be
\left|-\mg(z)d+\frac{1-d}{z}-1\right|\geq  \frac12,
\ee
which implies
\be
|1-z -d -z \,d\, m_N (z)|\geq \frac12|z|.
\ee
Insert it into \eqref{IDMzs}, we obtain when $z=P_{10}$,
\be\label{temp1.18}
\left|\mg(z)+\frac{1}{z  - (1-d) + z\, d \,\mg (z)}\right|\leq C\delta.
\ee
Denote the solutions of
\be\label{defeqS}
S+\frac{1}{z  - (1-d) + z\, d \,S}=\Delta
\ee
by $S^\Delta_\pm(z)$. Explicit calculation shows
\be\label{EExpl}
S^\Delta_\pm(z)= \frac{1-d-z \pm i (1+d\Delta)
\sqrt{(\lambda^\Delta_+-z)(z-\lambda^\Delta_-)} }{2dz} + \frac{\Delta}2,
\ee
where
\be
\lambda^\Delta_\pm\equiv\left(\frac{\sqrt{1+\Delta(d-d^2)}\pm \sqrt{d}}{1+\Delta d}\right)^2.
\ee
With the  notations:
\be
S_\pm(z)\equiv S_\pm^0(z),
\ee
we note $m_W(z)=S_+(z)$ (see \eqref{defmW}). Then the following lemma implies, with
${\rm Im}(m_N(z))> 0$, ${\rm Im}S_+(P_{10})>0$, ${\rm Im}S_-(P_{10})<0$,
that  \eqref{resultmNmWonline} holds for  $z_0=P_{10}$ if $\delta$ is small enough.
\begin{lemma}\label{SPM} Let $S^\Delta_\pm(z)$ be the solutions of \eqref{defeqS}.
Let $z=E+i\eta$ and $\frac12\lambda_-\leq E\leq 10$. For sufficiently small $\Delta$,
depending on $d$,
\be\label{resultSpm}
\max\Big\{|S^\Delta_+(z)-S_+(z)|,\,\,|S^\Delta_-(z)-S_-(z)|\Big\}\leq
C\frac{\Delta}{\sqrt{\kappa(E)+\Delta}}.
\ee
\end{lemma}
{\it Proof of Lemma \ref{SPM}.} First, when $\Delta$ is small enough, an easy calculation shows
\be\label{Wdede}
\widetilde \Delta\equiv\max_{\pm}\left\{|\lambda_\pm^\Delta-\lambda_\pm|\right\}\leq C|\Delta|.
\ee
Therefore, we have
\be
\max\Big\{|S^\Delta_+(z)-S_+(z)|,\,\,|S^\Delta_-(z)-S_-(z)|\Big\}\leq C|\Delta|+C\left|
\sqrt{(\lambda^\Delta_+-z)(z-\lambda^\Delta_-)}-\sqrt{(\lambda_+-z)(z-\lambda_-)}\right|.
\ee
Let $a=(z-\lambda_-)(\lambda_+-z)$ and $b=(z-\lambda^\Delta_-)(\lambda^\Delta_+-z)-
(z-\lambda_-)(\lambda_+-z)$. Note that $|b| \leq C \wt\Delta$ and therefore, by \eqref{Wdede},
$|b| \leq C \Delta$. Hence, \eqref{resultSpm} follows from
$|(\lambda_+-z)(z-\lambda_-)|\geq C\kappa(E)$ and from the inequality
\be
\left|\sqrt{a+b}-\sqrt a\right|\leq C\frac{|b|}{\sqrt{|a|+|b|}}.
\ee
which holds for any complex number $a$ and $b$.
\qed
\bigskip
\par Now we prove \eqref{resultmNmWonline}
for  the case $z_0\neq P_{10}$.  We first note that the two solutions of
\eqref{IDMzs} are  $S_\pm(z)$ when $Y^{(j)}=0$. One can check that for $z\in L(z_0,P_{10})$,
these two solutions are bounded by some constant $C_1$:
\be\label{temp6.20}
|S_\pm(z)|\leq C_1,\quad  \mbox{i.e.},\,\,\, |z-(1-d)+z dS_\pm(z)|\geq C_1^{-1},
\ee
and
\be\label{temp6.21}
|S_-(z)-S_+(z)|\geq C\sqrt{\kappa(\mbox{Re}\, z)+\mbox{Im}\, z}.
\ee
Furthermore, $|S_-(z_0)-S_+(z_0)|$ can be bounded by $|S_-(z)-S_+(z)|$
for any  $z\in L(z_0,P_{10})$ as follows,
\be\label{temp7.33}
|S_-(z_0)-S_+(z_0)|\leq C\min_{z\in L(z_0,P_{10})}|S_-(z)-S_+(z)|.
\ee
\par On the other hand, for any $z\in L(z_0,P_{10})$,  we claim that if $\mg(z)$ is close
to $S_-(z)$ or $S_+(z)$, then it should be really close to $S_-(z)$ or $S_+(z)$, i.e., if
\be\label{temp6.23}
\min\{|\mg(z)-S_-(z)|,\,\,\, |\mg(z)-S_+(z)|\} \leq C_1^{-1}/20,
\ee
then
\be\label{temp6.22}
\min\{|\mg(z)-S_-(z)|,\,\,\, |\mg(z)-S_+(z)|\} \leq
C \frac{\delta }{\sqrt{\kappa(\mbox{Re}\, z)+\delta}}.
\ee

To see this, note that \eqref{temp6.23} together with \eqref{temp6.20} imply
\be
|z-(1-d)+z dm_N(z)|\geq \frac12C_1^{-1}.
\ee
Then with  \eqref{IDMzs} and \eqref{temp6.14}, we obtain that \eqref{temp1.18} holds for
any $z\in L(z_0,P_{10})$. Using Lemma \ref{SPM} again, we have \eqref{temp6.22}.
\par We have seen that \eqref{resultmNmWonline}  and \eqref{temp6.23} (for small $\delta$)
hold when $z=P_{10}$. Because $\mg(z)$, $S_\pm(z)$ are  continuous functions of $z$, with
\eqref{temp6.22} and \eqref{temp6.23},  we can see that when $\delta$ is small enough,
\eqref{temp6.22} holds for every $z\in L(z_0, P_{10})$.  This result shows that $\mg(z)$
must be close to at least one of $S_+(z)$ and $S_-(z)$ and it is close to $S_+(z)$ when $z=P_{10}$.
\par Now we claim that if  $\mg(z_0)$ were close to $S_-(z_0)$, i.e.,
\be\label{mgz0s-}
|\mg(z_0)-S_-(z_0)| \leq C \frac{\delta }{\sqrt{\kappa(E)+\delta}},
\ee
then $\mg(z_0)$ is also close to $S_+(z_0)$, which implies that $\mg(z_0)$ is
always close to  $S_+(z_0)$.
\par
Again, with the continuity of $\mg(z)$ and $S_\pm(z)$ and \eqref{temp6.22},  if $\mg(z_0)$ is
close to $S_-(z_0)$ in the sense of \eqref{mgz0s-}, then there exists $z\in L(z_0,P_{10})$
such that $m_N(z)$ is close to both of $S_-(z)$ and $S_+(z)$, i.e.,
\be
|S_+(z)-S_-(z)|\leq 2C \frac{\delta }{\sqrt{\kappa(\mbox{Re}\, z)+\delta}},
\ee
which implies
\be\label{temp7.39}
|S_+(z)-S_-(z)|\leq 2C \frac{\delta }{\sqrt{C\kappa(E)+\delta}}.
\ee
Combining \eqref{temp7.39} and \eqref{temp7.33}, we obtain
\be
|S_+(z_0)-S_-(z_0)|\leq C \frac{\delta }{\sqrt{\kappa(E)+\delta}}.
\ee
Together with \eqref{mgz0s-}, we obtain  $m_N(z_0)$ is  still close to the $S_+(z_0)$.
It means that \eqref{resultmNmWonline} holds for all $z_0$'s in our assumption,
using the fact $S_+(z_0)=\mW(z_0)$.
This completes the proof of  Lemma \ref{mNmWonline}.
\qed
\bigskip

The following lemma shows that the expectation value of $\mg(z)$ is close to $m_W(z)$.

\begin{lemma}\label{EMg2}
Let $z=E+i\eta$, such that $N^{-1+\eps}\leq \eta\leq \frac12E$, $\frac12\lambda_-\leq E\leq 10$,
for some $\e>0$. Then we have
\be\label{resEmN2}
\left|\E \mg (z)- \mW(z)\right|\leq \frac{C}{N\eta \kappa^{3/2}},
\qquad \kappa=|(\lambda_+-E)(E-\lambda_-)|,
\ee
for large enough $N$ depending on $\e$.
\end{lemma}
\par {\it Proof of Lemma \ref{EMg2}.}
Using \eqref{temp6.20} and the estimate \eqref{PmNmW} from Proposition \ref{mainprop}, we
have
\be\label{EmgmWC}
|\E\, \mg(z)|\leq C,\,\,\, |\mW(z)|\leq C
\ee
uniformly in $z=E+i\eta$  within the range $N^{-1+\eps}\leq \eta\leq \frac12E$,
$\frac12\lambda_-\leq E\leq 10$.

We can  assume that $N\eta \kappa^{3/2}$ is much greater than 1, otherwise \eqref{resEmN2} is trivial.
Combining $N\eta \kappa^{3/2}>1$ and $N\eta>N^{\e}$, we obtain $N\eta \kappa\geq N^{\e/3}$.
Using  \eqref{IDMz}, we write $\E m_N(z)$ as
\be\label{IDMz2}
\E \mg(z)=-\frac{1}N\E\left(\sum_{j=1}^N\frac1{B-zd(\mg(z)-\E\, \mg(z))+Y^{(j)}(z)}\right),
\ee
where $B\equiv z-(1-d)-z\,d\, \E\mg(z)$.
Then, with \eqref{PmNmW}, we know
\be\label{temp7.30}
\E\left|\mg-\E \mg\right|^2\leq O\left( \frac{1}{N\eta \kappa}\right).
\ee
{F}rom \eqref{defXj}, \eqref{EXj0}, \eqref{defYj}, \eqref{eq:err2} and \eqref{PYJ}, we obtain
\be\label{temp7.31}
\E Y^{(j)}(z)=\E\left(\frac{1}{M} - \frac{z}{M} \left( (N-1) m_{N-1}^{(j)} (z) - N\mg (z)
\right)\right)=O\left(\frac{1}{N\eta}\right),
\ee
and
\be\label{temp7.32}
\E\left|Y^{(j)}\right|^2\leq O\left(\frac{1}{N\eta}\right).
\ee
Using  \eqref{PmNmW}, \eqref{temp6.20} and $N\eta\kappa\geq N^{\e/3}$, we obtain that $|B|$
is bounded from below by a constant $C_0$.
Furthermore,
for some $\delta>0$,
\be\label{conditionBzd}
\P\left(|B-zd(\mg(z)-\E(\mg(z)))+Y^{(j)}(z)|\leq C_0/2\right)\leq Ce^{-cN^\delta}.
\ee
Denote $a=-zd(\mg(z)-\E\mg(z))+Y^{(j)}(z)$, then
\be\label{E'b-a}
\E'\left(\frac{1}{B+a}\right)=\frac{1}{B}-\E'(a)B^{-2}+O(\E' a^2)|B|^{-3},
\ee
where $\E'$ is the conditional expectation  under the condition:
$|B-zd(\mg(z)-\E\mg(z))+Y^{(j)}(z)|\geq C_0/2$. Because $m_N(z)$ and $Y^{(j)}(z)$
are bounded from above by a polynomial of $M$, inserting 
\eqref{conditionBzd}  into \eqref{E'b-a}, we obtain
\be
\left|\E \mg(z)+\frac{1}{z-(1-d)+zd\E \mg(z)}\right|\leq C|\E(a)|+C\E a^2.
\ee
Combining this with  \eqref{temp7.30}, \eqref{temp7.31} and \eqref{temp7.32}, we obtain:
\be
\left|\E \mg(z)+\frac{1}{z-(1-d)+zd\E \mg(z)}\right|\leq \frac{C}{N\eta \kappa}.
\ee
Using Lemma \ref{SPM},    we have
\be
\min\bigg\{|\E \mg(z)-S_+(z)|,\,\,|\E \mg(z)-S_-(z)|\bigg\}\leq \frac{C}{N\eta \kappa^{3/2}}.
\ee
Using this inequality, and $S_+(z)=\mW(z)$, we can easily obtain \eqref{resEmN2}
for $z=P_{10}$. Consider now $z=E+i\eta\neq P_{10}$. If $\E \mg(z)$ is closer
to $S_-(z)$ than $C(N\eta \kappa^{3/2})^{-1}$, then
by the continuity of $\E\mg(z)$,  there exists $z'\in L(z,P_{10})$,
such that, $\E \mg(z')$ is close to both of $S_+(z')$ and $S_-(z')$, i.e.,
\be
|S_+(z')-S_-(z')|\leq \frac{C}{N{ \rm Im} \,z'
 [\kappa({ \rm Re} z')]^{3/2}}\leq
\frac{C}{N\eta \kappa^{3/2}}.
\ee
Together with \eqref{temp7.33}, we obtain that $\E \mg(z)$ is also  close
to $S_+(z)=\mW(z)$ and complete the proof.
\qed
\bigskip
\par Now we give an alternative bound on $\E\mg(z)$.
\begin{lemma}\label{EMg1}
Let $z=E+\eta i$, $N^{-1+\eps}\leq \eta\leq E/2$, $\lambda_-/2\leq E\leq 10$ and $\eps>0$.
Suppose $N\kappa\eta\geq N^{\e'}$ for some $\e'>0$, we have
\be\label{resEmN1}
\left|\E \mg(z)- \mW(z)\right|\leq \frac{C}{N\eta^{3/2}\kappa^{1/2}},
\ee
when $N$ is sufficiently large (depending on $\e'$).
\end{lemma}
\par {\it Proof of Lemma \ref{EMg1}.} We only prove the  case of the real sample covariance matrix.
The case of the complex sample covariance matrix can be treated similarly.
\par First, we show
\be
\label{Neta32}\E|\mg(z)-\E\mg(z)|\leq\frac{C}{N\eta^{3/2}}.
\ee
Let $\lambda_\alpha$ and $u_\alpha$ be the eigenvalues and eigenvectors of $H=A^*A$.
The derivative of  $\lambda_\alpha$ with respect to the $(i,j)$-th matrix
element $A_{ij}$ is given by
\be
\frac{\partial \lambda_\alpha}{\partial A_{ij}}=2(A u_\alpha )(i)u_\alpha(j).
\ee
Using $\sum_{j}u_{\alpha}(j)u_{\beta}(j)=\delta_{\alpha,\beta}$ and
$\sum_{i} (A u_{\alpha})(i)(Au_{\beta})(i)=\lambda_{\alpha}\delta_{\alpha,\beta}$,
one can obtain the following result, as in (3.3) of \cite{ESY1},
\be
\E|\mg(z)-\E\mg(z)|^2\leq\frac{C}{N^4}\E\left(\sum_{\alpha}
\frac{\lambda_\alpha}{|\lambda_\alpha-z|^4}\right).
\ee
Then with  Lemma \ref{lm:up}, as in (3.6) of \cite{ESY1}, we obtain \eqref{Neta32}.
\par Let $B\equiv z-(1-d)+z\,d\, \E\mg(z)$, $a_1=zd(\mg(z)-\E\mg(z))$ and $a_2=Y_j$ for each $j$.
Using the assumption $N\eta\kappa\geq N^{\e'}$ for some $\e'$, we obtain that $|B|$ is bounded
from below by a constant $C_0$  and for some $\delta>0$,
\be\label{Ba1a2}
\P\left(|B-a_1-a_2|\leq C_0/2\right)\leq e^{-N^\delta}.
\ee
We have
\be
\E'\frac{1}{B-a_1-a_2}=\frac{1}{B}+O( B^{-2}(\E'|a_1|))+
\frac{1}{B^2}\E'(a_2)+O(B^{-3}\E'(a_2^2)),
\ee
where $\E'$ is the conditional expectation  under the condition:
$|B-zd(\mg(z)-\E\mg(z))+Y^{(j)}(z)|\ge C_0/2$. Combining  this
with \eqref{Ba1a2}, \eqref{Neta32}, \eqref{temp7.31} and \eqref{temp7.32}, we obtain
\be\label{EmgNe32}
\left|\E \mg(z)+\frac{1}{z-(1-d)+z\,d\, \E\mg(z)}\right|\leq \frac{C}{N\eta^{3/2}}.
\ee
As in the proof of Lemma \ref{EMg2}, with  a continuity argument and Lemma
\ref{SPM}, we can obtain \eqref{resEmN1} from \eqref{EmgNe32} and complete the proof.
\qed

\section{Verifying Assumption III}\label{sec:check}

The following theorem 
gives the estimate \eqref{assum2} for the singular
values of the sample covariance matrices.

\begin{theorem}\label{maintheoremWish}  Assume that the single site distribution $\rd\nu$
of the entries of $\sqrt{M}A$ satisfies the logarithmic Sobolev inequality \eqref{LSI}.
Recall that $\rho_W$ in \eqref{defrhow}
denotes the density in Marchenko-Pastur law. Define
$\gamma_j\in [\lambda_-,\lambda_+]$, $(j=1,2,\ldots,N)$ with the relation
\be
\int_{-\infty}^{\gamma_j}\rho_W(x)dx=\frac{j}{N}.
\ee
Denote $x_j=\lambda_j^{1/2}$ the singular values of $A$. Then there exists
 $\delta>0$, such that,
\be\label{mainxga}
\frac{1}{ N}\E\sum_{j=1}^N\left|x_j-\gamma_j^{1/2}\right|^2\leq CN^{-1-\delta}.
\ee
\end{theorem}

{\bf Remark.} An analogous result holds for the eigenvalues of the Wigner matrices; 
the proof is similar and we will not give the details here.
We only point out that the key ingredients of the argument
below are: (i) apriori bound on the extreme eigenvalues (see Lemma \ref{boundKK} and
the remark afterwards); (ii) concentration of the local density of states
(used in Lemma \ref{lemma67}). We also critically use the
fact that the density of states (semicircle law or Marchenko-Pastur
law) has a square root singularity at the edges.

The local semicircle law needed in the analogue of Lemma \ref{lemma67} 
for Wigner matrices has been proven 
for hermitian matrices (see \cite{ERSY} and references therein)
but the proof is  valid for symmetric and quaternion self-dual 
matrices as well. The extension to symmetric matrices
is trivial. For the case of quaternion self-dual  matrices,
the only additional observation  is that the non-commutativity of the
quaternions is irrelevant in the arguments because the common starting point
of our papers \cite{ERSY, ESY1, ESY2, ESY3, ESY4} is an identity on 
the diagonal elements of the Green's function that involves only 
complex numbers. For simplicity, we present it only for
the (1,1) diagonal element $G_z(1,1)$ of $(H-z)^{-1}$, where
$H$ is an $N\times N$ quaternion self-dual matrix and $z\in\C$:
\be
   G_z(1,1)=   \frac{1}{h -z -  \ba^+ \cdot (B-z)^{-1}\ba}
 \cong \Big[ h-z - \frac{1}{N}
\sum_{\al=1}^{N-1}\frac{\xi_\al}{\lambda_\al- z}\Big]^{-1},
\label{G11}
\ee
in particular, $G_z(1,1)$ is a diagonal quaternion 
thus it can be identified with a complex number
via the identification \eqref{idd}.
Here $h\in \R$, $\ba \in \bH^{N-1}$ and $B$ is an $(N-1)\times (N-1)$
quaternion self-dual matrix obtained from the following decomposition of $H$
\be\label{Hd} H = \begin{pmatrix} h & \ba^+ \\
\ba & B
\end{pmatrix}.
\ee
The real numbers  $\lambda_1\leq \lambda_2 \leq
\ldots \leq \lambda_{N-1}$ denote the eigenvalues of $B$
and the nonnegative real numbers $\xi_\al$ are given by
$$
\xi_{\alpha} = {N} (\ba^+ \cdot \bu_{\alpha})( \bu_{\alpha}^+ \cdot   \ba)
= N |\ba^+ \cdot \bu_{\alpha}|^2
$$
where  $\bu_{\alpha}\in \bH^{N-1}$ is the normalized
eigenvector of $B$ associated with the eigenvalue
$\lambda_{\alpha}$. The dot product of two quaternionic
vectors, $\ba, {\bf b}\in \bH^{N-1}$ is defined as
$$
    \ba^+\cdot {\bf b} := \sum_{n=1}^{N-1} a_n^+ b_n.
$$
The proof of \eqref{G11} is a straightforward computation.
This identity is the key to extend our results on 
the local semicircle law for quaternion self-dual matrices
without any further modifications.

\bigskip

\par Now we return to the the proof of Theorem \ref{maintheoremWish}
and  we start with some preparatory lemmas.
First, we recall the following result from \cite{FSo}.
\begin{lemma}\label{boundKK}(Corollary V.2.1 of \cite{FSo})
Define
\be n^\lambda(E) \equiv \frac1N \E \left[\#\{\lambda_j\leq E\}\right],
\ee
then $n^\lambda(\lambda_--N^{-1/5})\leq Ce^{-N^\e}$ and $1-n^\lambda(\lambda_++N^{-1/5})
\leq Ce^{-N^\e}$ for some $\e>0$. Therefore for any $1\leq j\leq N$,
\be\label{aledge}
\lambda_--CN^{-1/5}\leq \E\lambda_j\leq \lambda_++CN^{-1/5}.
\ee 
\end{lemma}

{\bf Remark.}
In fact,  the error term in \cite{FSo} is $N^{-2/3+\e}$ instead of $N^{-1/5}$ but we 
will use only the weaker bound \eqref{aledge} in order to indicate that
our proof goes through for Wigner matrices with not necessarily symmetric
distributions as well. In the latter case only $N^{-1/4+\e}$ has been proven
by Vu in \cite{Vu} for compactly supported distribution
$\nu$ with an effective dependence of the constant $C$ on
the support. This effective dependence is necessary to remove
the compact support condition as in Lemma 4.1 of \cite{ESY4}.
Strictly speaking, the result in \cite{Vu} was stated only for symmetric
Wigner matrices but it holds for hermitian and quaternion self-dual 
Wigner matrices as well, since the key estimate (equation (5) in \cite{Vu})
is independent of the matrix ensemble.

\begin{lemma}\label{lemma67}
Recall that $\rho_W$ in \eqref{defrhow} denotes the density  in Marchenko-Pastur law. Let
\be
n^\lambda _W(E) =\int_{-\infty}^{E}\rho_W(x)dx,
\ee
then
\be\label{nlambda67}
\int_0^\infty \left|n^\lambda(E)-n_W^\lambda(E)\right|dE\leq C N^{-6/7},
\ee
and
\be\label{supnlambda37}
\sup_{E}\left|n^\lambda(E)-n_W^\lambda(E)\right|\leq CN^{-3/7}.
\ee
\end{lemma}
{\it Proof of Lemma \ref{lemma67}.}
To prove \eqref{nlambda67}, with Lemma \ref{boundKK}, one only needs to prove
\be\label{temp1.35}
\int_{\frac12\lambda_-}^{2\lambda_+} \left||n^\lambda(E)-n^\lambda(\lambda_-/2)|-n_W^\lambda(E)
\right|\rd E\leq C N^{-6/7}.
\ee
To this end, we first note that $|\E\, \mg(z)|$ is bounded uniformly
for $z=E+i\eta$, such that $N^{-1+\e}\leq \eta\leq E/2$ and $\frac{1}{2}\lambda_-\le
E\leq 10$ (see \eqref{EmgmWC}). Moreover, by \eqref{resEmN1} and \eqref{resEmN2},
the conditions of Lemma B.1 in \cite{ERSY} are satisfied and thus we obtain \eqref{temp1.35}.
Following the proof of Lemma B.1 in \cite{ERSY}, we see that
this lemma is  still valid  if $(\log N)^4$ in (B.2),
(B.3) and (B.4) is replaced with $N^{\e}$ for small enough $\eps>0$.
\par To prove \eqref{supnlambda37}, for fixed $E$, we can assume $n^\lambda(E)>n_W^\lambda(E)$
and denote $\Delta=n^\lambda(E)-n_W^\lambda(E)$. Because $n^\lambda(E)$ is an increasing
function and the derivative of $n_W^\lambda(E)$ is bounded by
$\|\rho_W\|_\infty=\frac1\pi(d-d^2)^{-1/2}$, we have:
\be
n^\lambda(E') - n_W^\lambda(E') \geq \Delta-C(E'-E)>0, 
\,\,\,{\rm when}\,\,\,E\leq E'\leq E+C^{-1}\Delta.
\ee
Integrating both sides, we obtain
\be
\int_{E}^{E+C^{-1}\Delta} |n^\lambda(E')- n_W^\lambda(E')| \rd E'\geq O(\Delta^2).
\ee
Using \eqref{nlambda67}, it follows that $\Delta\leq O(N^{-3/7})$.
\qed
\bigskip

\par Similarly to the calculation in Theorem 3.1 of \cite{ESY1},
by using the logarithmic Sobolev inequality, we
have
\begin{lemma}For $j, K\in \mathbb N$, such that, $j+K\leq N+1$, let
$\lambda_{j,K}=K^{-1}\sum_{i=0}^{K-1}\lambda_{j+i}$, then for any $\delta>0$
\be\label{rangexjk}
\mathbb P\left(|\lambda_{j,K}-\E(\lambda_{j,K})|\geq N^{-1/2+\delta}K^{-1/2}\right)\leq Ce^{-N^{\delta/2}},
\ee
with $C$ depending on $\delta$.
\end{lemma}
\qed
\bigskip
\par We will say that an event $A$, depending on $N$, occurs with an
{\it extremely high probability} if $\P(A)\geq 1-N^{-C}$ for any $C$ and sufficiently large $N$.
\begin{lemma}\label{averagealga}
\par Recall that $\alpha_j$ is defined as $\E(\lambda_j)$. Suppose that there exist
sufficiently small positive numbers $\e_1$, $\e_2$ and $\e_3$, such that, 
\be\label{Wassum1}
\lambda_-+N^{-2\e_2}\leq \lambda_{j_-}\leq \lambda_-+N^{-\e_2}\mbox{,\,\,\,}\lambda_+-N^{-2\e_2}\geq
\lambda_{j_+}\geq \lambda_+-N^{-\e_2}
\ee
and that
\be\label{Wassum2}
|\lambda_j-\alpha_j|\leq N^{-\frac12-\e_3}\,\,\,\,{\rm for\,\,\,} j_-< j< j_+,
\ee
hold with an extremely high probability, where we introduced the notations $j_-\equiv N^{1-\e_1}$
and $j_+\equiv N-N^{1-\e_1}$.
Then for some $\e>0$, we have
\be\label{algamresult}
\frac{1}{N}\sum_{j}|\alpha_j-\gamma_j|^2\leq N^{-1-\e}.
\ee
\end{lemma}

{\it Proof.}
By symmetry, we only need to prove that \eqref{algamresult} holds for the sum on
the indices $j$ with $ \gamma_j\leq\alpha_j$.
Introduce the notation
\be
n^\alpha(E): =\frac1N \#[ \{\alpha_j\leq E\}].
\ee
The estimate  \eqref{rangexjk}, with $K=1$, implies that $\max_j|\lambda_j-\alpha_j|
\leq N^{-1/2+\delta}$ holds with an extremely high probability, for any positive
$\delta$. Therefore, we can bound $n^\alpha(E)$ from above by (for any $E$)
\be
n^\lambda(E+N^{-1/2+\delta})=\frac1N \E[ \#\{\lambda_j\leq E+N^{-1/2+\delta}\}]\geq
\frac1N \#[ \{\alpha_j\leq E\}]-CN^{-100}=n^\alpha(E)-CN^{-100}.
\ee
Similarly, we can obtain the lower bound. Putting them together, we have that:
\be\label{bdnalE1}
CN^{-100}+n^\lambda(E+N^{-1/2+\delta})\geq n^\alpha(E) \geq n^\lambda(E-N^{-1/2+\delta})-CN^{-100}
\ee
holds for any $E$. The assumption \eqref{Wassum1} implies that
\be
\lambda_j\notin [\lambda_-+N^{-\e_2}, \lambda_+-N^{-\e_2}]
\ee
holds with an extremely high probability, for  any $j\leq j_-$ or $j\geq j_+$.
For the other $j$'s, for which $\lambda_j$ may appear in $[\lambda_-+N^{-\e_2}, \lambda_+-N^{-\e_2}]$,
we use \eqref{Wassum2} and  obtain the following improved bound on $n^\alpha(E)$:
when  $\lambda_-+N^{-\e_2}\leq E\leq \lambda_+-N^{-\e_2}$,
\be\label{bdnalE2}
CN^{-100}+n^\lambda(E+N^{-1/2-\e_3})\geq n^\alpha(E) \geq n^\lambda(E-N^{-1/2-\e_3})-CN^{-100}.
\ee
Let $F(E)$ be  a continuous and differentiable function, such that $N^{-1/2-\e_3}\leq F(E)\leq
N^{-1/2+\delta}$, for  $0\leq \delta\leq
\frac{1}{10}\min\{\e_1,\e_2,\e_3\}$, $F(E)=N^{-1/2-\e_3}$ for $\lambda_-+2N^{-\e_2}
\leq E\leq \lambda_+-2N^{-\e_2}$, $F(E)=   N^{-1/2+\delta}$ for $E\leq \lambda_-+N^{-\e_2}$
or $E\geq \lambda_+-N^{-\e_2}$ and $|F'(E)|\leq N^{-\delta}$. Combining \eqref{bdnalE2}
and \eqref{bdnalE1}, we  obtain
\be\label{bdnalE3}
CN^{-100}+n^\lambda(E+F(E))\geq n^\alpha(E) \geq n^\lambda(E-F(E))-CN^{-100}.
\ee

On the other hand,  we have
\be
\alpha_j-\gamma_j= \int_{\R}{\bf 1}\left(n_W(E)\geq \frac jN> n^\alpha(E)\right)\rd E,
\ee
for any $j$, such that, $\alpha_j>\gamma_j$.
Therefore we  can write 
\begin{align}\label{temp7.60}
\frac{1}{N} & \sum_{j: \gamma_j\leq\alpha_j} |\alpha_j-\gamma_j|^2 \\
&=\frac2N\sum_{j}\int\int_{E'\leq E}{\bf 1}\left(n_W(E)\geq \frac jN> n^\alpha(E)\right)
{\bf 1}\left(n_W(E')\geq \frac jN> n^\alpha(E')\right)\rd E\rd E'\non \\
&=\frac2N\sum_{j}\int\int_{E'\leq E}{\bf 1}\left(n_W(E)\geq \frac jN>
n^\alpha(E) + C N^{-100} \right) {\bf 1}\left(n_W(E')\geq \frac jN> n^\alpha(E')\right)\rd E\rd E' ,
\non
\end{align}
where in the second line we used the fact that the difference between $j/N$ and
$n^{\alpha} (E)$ must be a multiple of $N^{-1}$. Since $\max_j \lambda_j\leq 10$
holds with an extremely high probability,  using
\eqref{bdnalE3}, we can replace $n^\alpha(E)$ with $n^\lambda(E-F(E))$ in \eqref{temp7.60}, i.e.,
\begin{align}\label{temp7.602}
\frac{1}{N}\sum_{j: \gamma_j\leq\alpha_j}&|\alpha_j-\gamma_j|^2-N^{-10}\\\nonumber
\leq &\,\,
\frac2N\sum_{j}\int\int_{E'\leq E}{\bf 1}\left(n_W(E)\geq \frac jN> n^\lambda(E-F(E))
\right){\bf 1}\left(n_W(E')\geq \frac jN> n^\alpha(E')\right)\rd E\rd E'.
\end{align}
Change  the variable from $E$ to $t(E)=E-F(E)$.  With $|F'|\leq N^{-\delta}$,
we obtain $F(t)=(1+O(N^{-\delta}))F(E)$ and $\rd t/\rd E=(1+O(N^{-\delta}))$. Thus
\begin{align}\label{temp7.59}
\frac{1}{N}\sum_{j: \gamma_j\leq\alpha_j}&|\alpha_j-\gamma_j|^2-N^{-10}\\\nonumber
\leq &\,\, \frac3N\sum_{j}\int\int_{E'\leq E(t)}{\bf 1}\left(n_W(t+2F(t))\geq
\frac jN> n^\lambda(t)\right){\bf 1}\left(n_W(E')\geq \frac jN> n^\alpha(E')\right)\rd t\rd E',
\end{align}
where $E(t)$ is the inverse function of $t(E)$.
Note, when ${\bf 1}\left( \cdots\right){\bf 1}\left( \cdots\right)=1$ in \eqref{temp7.59},
we have  $n_W(E')\geq j/N> n_\lambda(t)$. Define the inverse function of $n_W$ as
$n_W^{-1}(1)=\lambda_+$, $n_W^{-1}(0)=\lambda_-$ and $n_W^{-1}(n_W(x))=x$ for $0<n_W(x)<1$. Then
\be
t+2F(t)\geq E\geq E'\geq n_W^{-1}(n^\lambda(t)).
\ee
\par Inserting this inequality into \eqref{temp7.59} and performing the $dE'$
integration, we can see that
\begin{align}
\frac{1}{N}\sum_{j: \gamma_j\leq\alpha_j}|\alpha_j-\gamma_j|^2-\frac{1}{N^5}
&\leq \frac CN\sum_{j}\int{\bf 1}\left(n_W(t+2F(t))\geq
\frac jN> n^\lambda(t)\right)|t- n_W^{-1}(n^\lambda(t))+2F(t)|\rd t \non \\ 
&\leq C\int \left(|n_W(t+2F(t))-n^\lambda(t)|+N^{-1} \right)
\cdot|t- n_W^{-1}(n^\lambda(t))+2F(t)|\rd t \non \\  
&\leq C\left(A_1+A_2+A_3+A_4\right),\label{sum4A}
\end{align}
where we expanded \eqref{sum4A} into four terms:
\be
\begin{split}
A_1&= \int|n_W(t+2F(t))-n_W(t)|F(t)\rd t\\\nonumber
A_2&= \int (|n_W(t)-n^\lambda(t)|+N^{-1})F(t)\rd t\\\nonumber
A_3&=\int|n_W(t+2F(t))-n_W(t)|\cdot|t- n_W^{-1}(n^\lambda(t))|\rd t\\\nonumber
A_4&=\int (|n_W(t)-n^\lambda(t)| +N^{-1}) \cdot|t- n_W^{-1}(n^\lambda(t))|\rd t.
\end{split}
\ee
Since $n_W'(t)=\rho_W(t)\leq C$, $F(t)=N^{-1/2-\e_3}$ when $\lambda_-+2N^{-\e_2}\leq
t\leq \lambda_+-2N^{-\e_2}$ and $F(t)\leq N^{-1/2+\delta}$ for any $E$, we obtain
$A_1\leq N^{-1-\e}$, for some $\e>0$. Next, from \eqref{nlambda67} and
$F(t)\leq N^{-1/2+\delta}$ for any $t$, we can see  $A_2\leq (N^{-6/7}+N^{-1}
) N^{-1/2+\delta}\leq  N^{-1-\e}$.
\par To prove  $A_3\leq N^{-1-\e}$, we start with writing $A_3$ as
\be\label{reA3}
A_3=\Bigg[ \int_{\lambda_+<t} +\int_{ t<\lambda_-}+\int_{\lambda_-\leq
t\leq \lambda_-+E_1} + \int_{\lambda_+-E_1\leq t\leq \lambda_+} +
\int_{\lambda_-+E_1\leq t\leq \lambda_+-E_1}\Bigg] \Xi(t)\rd t,
\ee
where we set $E_1\equiv N^{-1/4}$ and  $$\Xi(t)\equiv |n_W(t+2F(t))-n_W(t)|
\cdot|t- n_W^{-1}(n^\lambda(t))|.$$
\par The first term on the r.h.s. of (\ref{reA3}) is equal to zero,
since $n_W$ is constant outside $[\lambda_-,\lambda_+]$. The second
 term can be bounded by $N^{-1-\e}$,
for some $\e>0$, using the facts $F(t)\leq N^{-1/2+\delta}$ and 
$n_W(\lambda_-+E)\leq CE^{3/2}$, i.e.,
\be
\int_{ t<\lambda_-}\Xi(t)\rd t\leq C\int_{\lambda_--N^{-1/2+\delta}}^{\lambda_-}
|F(t)|^{3/2}\rd t\leq N^{-1-\e}.
\ee
Now we prove that the third and fourth term of \eqref{reA3} are less
$N^{-1-\e}$, for some $\e>0$.

{F}rom the explicit definition of $n_W$, an easy calculation shows that, for
all $t \in (\lambda_-, \lambda_+)$, \begin{equation}\label{nW-1}
|t - n_W^{-1} (s)| \leq C |n_W(t) - s|^{2/3}
\end{equation}
which in particular implies that
\be\label{maxtnw}
\max_{t\in(\lambda_-,\lambda_+)}\left(\max_{|s-n_W(t)|
\leq CN^{-3/7}}|t- n_W^{-1}(s)|\right)\leq CN^{-2/7}.
\ee
Combining this with the fact $|n_W(t+2F(t))-n_W(t)|\leq C\|n'_W\|_\infty F(t)\leq
C N^{-1/2+\delta}$, we obtain that the third and fourth terms of \eqref{reA3}
are  less than $CN^{-2/7}N^{-1/2+\delta}N^{-1/4}\leq N^{-1-\e}$, for some $\e>0$.
\par

To bound the last term of \eqref{reA3}, we use, once again the bound $|n_W(t+2F(t))-n_W(t)|
\leq C N^{-1/2+\delta}$. {F}rom (\ref{nW-1}), we find therefore that
\[  \begin{split} \int_{\lambda_-+E_1\leq t\leq \lambda_+-E_1} \Xi(t)\rd t &\leq C N^{-1/2+\delta}
\int |n_W(t) - n^\lambda (t)|^{2/3}\rd t \\&\leq C N^{-1/2+\delta}
\left(\int |n_W (t) - n^{\lambda} (t)| \rd t\right)^{2/3} \leq C N^{-1-\e} \end{split} \]

\par At last, we prove $A_4\leq N^{-1-\e}$. We rewrite $A_4$ as
\be
A_4=\int_{t\notin(\lambda_-,\lambda_+)}\Sigma(t)\rd t+\int_{t\in(\lambda_-,\lambda_+)}\Sigma(t)\rd t,
\ee
where $\Sigma(t)\equiv (|n_W(t)-n^\lambda(t)|+N^{-1}) \cdot|t- n_W^{-1}(n^\lambda(t))|$.
When $t\notin(\lambda_-,\lambda_+)$, from \eqref{supnlambda37} and \eqref{maxtnw}, one can see that
\be
|t- n_W^{-1}(n^\lambda(t))|\leq \max_{|s-n_W(t)|\leq CN^{-3/7}}|t- n_W^{-1}(s)|
\leq C|t-\lambda_-||t-\lambda_+|+CN^{-2/7}.
\ee
So we have
\be\label{temp7.73}
\int_{t\notin(\lambda_-,\lambda_+)}\Sigma(t)\rd t=C\int_{t\notin(\lambda_-,\lambda_+)}\rd t
\left(\left|n^\lambda(t)-n_W(t)\right|+N^{-1}\right)
 \left(|t-\lambda_-||t-\lambda_+|+N^{-2/7}\right).
\ee
Using Lemma \ref{boundKK}, we have
\begin{align}
\eqref{temp7.73}\leq & \; C\int_{\lambda_--N^{-1/5}}^{\lambda_-}\rd t
\left( \left|n^\lambda(t)-n_W(t)\right|+N^{-1}\right)
\left(|t-\lambda_-||t-\lambda_+|+N^{-2/7}\right)
\\
&+\int_{\lambda_+}^{\lambda_++N^{-1/5}}\rd t
\left( \left|n^\lambda(t)-n_W(t)\right|+N^{-1}\right)
\left(|t-\lambda_-||t-\lambda_+|+N^{-2/7}\right)
+N^{-10}. \non
\end{align}
Here we also used the fact that for large $t$, $|n^\lambda (t) - 1|$ decays exponentially
fast to zero (see, for example, Lemma 7.3 of \cite{ESY1}, which is stated for matrices
with complex entries, but can be trivially extended to the case of real entries).
Together with  \eqref{nlambda67}, we obtain that
\be
\eqref{temp7.73}\leq C (N^{-6/7} +N^{-1}) \left(N^{-1/5}+N^{-2/7}\right)+N^{-10}\leq N^{-1-\e},
\ee
for some $\e>0$.
\par When $t\in(\lambda_-,\lambda_+)$,  with \eqref{maxtnw} and \eqref{nlambda67}, we can see
\be\label{temp7.74}
\int_{t\in(\lambda_-,\lambda_+)}\Sigma(t)\rd t\leq CN^{-6/7}N^{-2/7}.
\ee
Combining \eqref{temp7.73} and \eqref{temp7.74}, we obtain $A_4\leq N^{-1-\e}$ for
some $\e>0$. Together with \eqref{sum4A}, this compeletes the proof of Lemma \ref{averagealga}.
\qed
\bigskip
\par Next, we show that the assumptions \eqref{Wassum1} and \eqref{Wassum2} in Lemma
\ref{averagealga} always hold. First we prove \eqref{Wassum1} in the next
Lemma \ref{Gassum1} with an analogous proof as Lemma \ref{averagealga}.
Then in Lemma \ref{Gassum2} we show that \eqref{Wassum2} holds when \eqref{Wassum1} holds.
\begin{lemma}\label{Gassum1}
There exist  small positive numbers $\e_1$ and $\e_2$, such that,
\be\label{RGassum1}
\lambda_-+N^{-2\e_2}\leq \lambda_{j_-}\leq \lambda_-+N^{-\e_2}\mbox{,\,\,\,}
\lambda_+-N^{-2\e_2}\geq \lambda_{j_+}\geq \lambda_+-N^{-\e_2}
\ee
hold with an extremely high probability, where we recall the notations
$j_-\equiv N^{1-\e_1}$ and $j_+\equiv N-N^{1-\e_1}$.
\end{lemma}
{\it Proof.}
As in \eqref{bdnalE1}, for any $E$, $\delta>0$ and sufficiently large $N$, we have
\be
CN^{-100}+n^\lambda(E+N^{-1/2+\delta})\geq n^\alpha(E) \geq n^\lambda(E-N^{-1/2+\delta})-CN^{-100}.
\ee
So without any other assumptions, one can obtain  \eqref{sum4A}, if we set
$F(E)\equiv N^{-1/2+\delta}$ instead of $F(E)$ defined in the proof of Lemma
\ref{averagealga}. With  a similar argument as in the proof of Lemma \ref{averagealga}
but with this redefined $F(E)$, we have
\be\label{temp7.102}
\frac{1}{N}\sum_{i}\left|\alpha_i-\gamma_i\right|^2\leq C N^{-1+C\delta}.
\ee
Then we claim that \eqref{temp7.102} implies that
\be\label{temp7.76}
\sup_{j}\left|\alpha_j-\gamma_j\right|\leq N^{-\frac{1}{10}}.
\ee
We prove this claim by contradiction; assume that
for some $j_0$ we have $\left|\alpha_{j_0}-\gamma_{j_0}\right|\geq N^{-\frac1{10}}$.
By symmetry we can assume that $j_0\le N/2$, the case $j_0\ge N/2$ is analogous.
We start with the case  $j_0\le N^{1/2}$.  Then $\gamma_{j_0}\le \lambda_-+CN^{-1/4}$
and in this case $\alpha_{j_0}$ must be
larger than $\gamma_{j_0}$, otherwise
$\al_{j_0}\le \gamma_{j_0}- N^{-\frac{1}{10}} \le \lambda_- -\frac{1}{2} N^{-\frac{1}{10}} $
would contradict to  $\alpha_{j_0}\in [\lambda_--CN^{-1/5}, \lambda_++CN^{-1/5}]$, see
\eqref{aledge}.  Using
\be
|\gamma_i-\gamma_j|\leq CN^{-2/3}|i-j|
\label{gam}
\ee
for any $i,j$ and that $\al_j$ is monotone, we  obtain that
$$
 \al_j -\gamma_j \ge \al_{j_0} - \gamma_{j_0} - CN^{-1/6} \ge \frac{1}{2} N^{-\frac{1}{10}}
$$
for any $j$ such that $j_0 \le j \le j_0 + N^{1/2}$. Then
\be
 \sum_{j=j_0}^{j_0+N^{1/2}} |\al_j-\gamma_j|^2 \ge cN^{\frac{1}{2} - \frac{1}{5}}
\label{ss}
\ee
with some positive $c>0$ which would contradict to \eqref{temp7.102}.
Now we consider the case $j_0\ge N^{1/2}$. The previous argument
remains unchanged if $\al_{j_0} > \gamma_{j_0}$. If $\al_{j_0} < \gamma_{j_0}$, then we
use
$$
  \al_j -\gamma_j \le \al_{j_0} - \gamma_{j_0} + CN^{-1/6} \le - \frac{1}{2} N^{-\frac{1}{10}}
$$
for any $j$ such that $j_0 - N^{1/2} \le j \le j_0 $ and we obtain
$$
 \sum_{j=j_0-N^{1/2}}^{j_0} |\al_j-\gamma_j|^2 \ge cN^{\frac{1}{2} - \frac{1}{5}},
$$
which again contradicts to \eqref{temp7.102}. This completes the proof of
\eqref{temp7.76}.

On the other hand, the estimate  \eqref{rangexjk}, with $K=1$, implies
$\max_j|\lambda_j-\alpha_j|\leq N^{-1/2+\delta}$ holds with an extremely high probability.
Combining \eqref{temp7.76} with this fact, we can see that for any small enough $\e_1$,
there exists  $\e_2$ such that \eqref{RGassum1} holds,
 which completes the proof of Lemma \ref{Gassum1}.

\qed

\bigskip
The next Lemma guarantees the  assumption \eqref{Wassum2} in 
 Lemma \ref{averagealga},  given  \eqref{Wassum1}.

\begin{lemma}\label{Gassum2}
If there exist sufficiently small positive numbers $\e_1$ and $\e_2$, such that
\be\label{AGassum2}
\lambda_-+N^{-2\e_2}\leq \lambda_{j_-}\leq \lambda_-+N^{-\e_2}\mbox{,\,\,\,}
\lambda_+-N^{-2\e_2}\geq \lambda_{j_+}\geq \lambda_+-N^{-\e_2},
\ee
holds with an extremely high probability, then there exists $\e_3>0$ such that,
\be\label{assumf2}
|\lambda_j-\alpha_j|\leq N^{-\frac12-\e_3},\,\,\,{\rm for\,\,\,} j_-< j< j_+,
\ee
holds with an extremely high probability, where we recall the notations
$j_-\equiv N^{1-\e_1}$ and $j_+\equiv N-N^{1-\e_1}$.

\end{lemma}
{\it Proof.} For simplicity, we only prove the case of $j\leq N/2$, the case $j>N/2$
is analogous. Using \eqref{rangexjk}, for any $N/2\geq j> j_-$, $\delta>0$, with $K=N^{1/4}$, we have
\be\label{temp7.107}
\P(|\lambda_{j,K}-\E(\lambda_{j,K})|\geq N^{-5/8+\delta})\leq C e^{-N^{\delta}}.
\ee
Now we claim that, for $K=N^{1/4}$, $j_-< j< j_+$,
\be\label{xjkxj}
|\lambda_{j,K}-\lambda_j|\leq N^{-5/8}
\ee
holds with an extremely high probability, which implies
\be\label{ExjEkxj}
|\E\lambda_{j,K}-\E\lambda_j|\leq CN^{-5/8}.
\ee
To see \eqref{xjkxj}, first notice that
\be\label{xj+kxj}
\P(|\lambda_{j,K}-\lambda_j|\geq N^{-5/8})\leq \P(\lambda_{j+K}-\lambda_j\geq N^{-5/8}).
\ee
Suppose now that $\lambda_{j+K}-\lambda_j\geq N^{-5/8}$.  With the assumption \eqref{AGassum2},
we have that, for $j_-< j< N/2$,
\be
\lambda_j\in \left(\lambda_-+N^{-2\e_2}, \lambda_+-N^{-2\e_2}\right).
\ee
with an extremely high probability. Divide this interval into small intervals with
the length $\frac12N^{-5/8}$. By the local Marchenko-Pastur law, i.e., Corollary
\ref{localMPlaw}, the event that the  number of the eigenvalues in each piece
is larger than $CN^{1-3\e_2-5/8}$ holds with an extremely high probability.
On the other hand, if $\lambda_{j+K}-\lambda_j\geq N^{-5/8}$, then the total
number of eigenvalues in at least one of these intervals is less than $K=N^{1/4}$,
which implies that $\lambda_{j+K}-\lambda_j\leq N^{-5/8}$ holds with an extremely
high probability. Together with \eqref{xj+kxj}, we have  \eqref{xjkxj}.
Then combining \eqref{xjkxj}, \eqref{ExjEkxj} and \eqref{temp7.107}, we obtain
\eqref{assumf2} and complete the proof.
\qed
\bigskip
\par Now we are ready to prove  Theorem \ref{maintheoremWish}.

\medskip

{\it Proof of Theorem \ref{maintheoremWish}.} Note that the assumptions in Lemma
\ref{averagealga} are proved in Lemma \ref{Gassum1} and  \ref{Gassum2}.
Combining Lemma \ref{averagealga}, \ref{Gassum1} and  \ref{Gassum2},
we obtain \eqref{algamresult}, i.e.,
\be\label{algamresult2}
\frac{1}{N}\sum_{j}|\alpha_j-\gamma_j|^2\leq N^{-1-\e},
\ee
for some constant $\e>0$, where $\alpha_j $ is defined as $\E\lambda_j$.
Then we claim that for some constant $\e>0$,
\be\label{laalN-1}
\frac{1}{N}\sum_{j}|\lambda_j-\alpha_j|^2\leq N^{-1-\e}
\ee
holds with an extremely high probability. To see \eqref{laalN-1},
first notice that \eqref{rangexjk}, with $K=1$, implies that, for any $\delta$ and $j$
\be\label{temp7.113}
|\lambda_j-\alpha_j|\leq N^{-1/2+\delta}
\ee
holds with an extremely high probability. The estimate \eqref{assumf2}
shows that there exist $\e_1>0$ and $\e_3>0$ such that
\be\label{temp7.114}
|\lambda_j-\alpha_j|\leq N^{-\frac12-\e_3},\,\,\,{\rm for\,\,\,} N^{\e_1}< j< N-N^{\e_1}
\ee
holds with an extremely high probability. Combining this with \eqref{temp7.113}
for the remaining indices $j\leq N^{\e_1}$ or $j\geq N-N^{\e_1}$, we obtain \eqref{laalN-1}.
Together with \eqref{algamresult2}, we have:
\be\label{laga-1}
\frac{1}{N}\E\sum_{j}|\lambda_j-\gamma_j|^2\leq N^{-1-\e},
\ee
for some $\e>0$. Using the definition $x_j=\lambda_j^{1/2}$, one has
\be\label{xjgj}
|x_j-\gamma_j^{1/2}|=|\lambda_j-\gamma_j|(x_j+\gamma_j^{1/2})^{-1}\leq C |\lambda_j-\gamma_j|.
\ee
Inserting \eqref{xjgj} into \eqref{laga-1}, we obtain  \eqref{mainxga}
and complete the proof of Theorem \ref{maintheoremWish}.
\qed

\appendix

\section{Existence and restriction of the dynamics}
\label{sec:defdyn}

As in Section \ref{sec:flow},
we consider the Euclidean space $\bR^N$ with the normalized
measure $\mu = \exp(-N\cH)/Z$. The Hamiltonian $\cH$ is of the form \eqref{ham}
or \eqref{hamW}, for definiteness we discuss the first case,
the second case is fully analogous.
$\cH$ is symmetric
with respect to the permutation of the variables $\bx = (x_1, \ldots, x_N)$,
thus the measure can be restricted to the subset $\Sigma_N \subset\bR^N$
defined in \eqref{def:Sigma}. In this appendix we outline how to
define the dynamics \eqref{dy} with its generator, formally given by
$L= \frac{1}{2N}\Delta - \frac{1}{2}(\nabla \cH)\nabla$,
on $\Sigma_N$. The condition $\beta\ge 1$ and the specific
factors $\prod_{i<j}|x_j-x_i|^\beta$ will play a key role in the argument,
in particular, we will see that $\beta=1$ is the critical threshold
for this method to work.

We first recall the standard definition of the dynamics on $\bR^N$.
The quadratic form
$$
\cE(u,v) := \int_{\bR^N} \nabla u\cdot \nabla v \; \rd\mu
$$
is a closable Markovian symmetric form on $L^2(\bR^N,\rd\mu)$ with a domain
$C_0^\infty(\bR^N)$ (see Example 1.2.1 and Theorem 3.1.3 of \cite{Fu}).
This form can be closed with a form domain
$H^1(\bR^N, \rd\mu)$ defined as the closure of $C_0^\infty$
in the norm $\|\cdot \|_+^2 = \cE(\cdot, \cdot) + \|\cdot\|_2^2$.
The closure is called the Dirichlet form.
It generates a strongly continuous Markovian semigroup $T_t$, $t>0$,
on $L^2$ (Theorem 1.4.1 \cite{Fu}) and it can be extended to a contraction
semigroup to $L^1(\bR^N,\rd\mu)$, $\| T_tf\|_1\le \|f\|_1$ (Section 1.5 \cite{Fu}).
The generator $L$ of the semigroup, is defined via the Friedrichs extension
(Theorem 1.3.1 \cite{Fu}) and it is a positive self-adjoint operator
on its natural domain $D(L)$ with $C_0^\infty$ being the core. The generator
is given by $L= \frac{1}{2N}\Delta - \frac{1}{2}(\nabla \cH)\nabla$
on its domain (Corollary 1.3.1 \cite{Fu}). By the spectral theorem,
$T_t$ maps $L^2$ into $D(L)$, thus with the notation $f_t = T_t f$ for
some $f\in L^2$, it holds that
$$
 \pt_t f_t = Lf_t, \qquad t>0, \quad \mbox{and}\quad \lim_{t\to0+0}\| f_t-f\|_2 = 0.
$$
Moreover,  by approximating
$f$ by $L^2$ functions and using that $T_t$ is contraction in $L^1$
(Section 1.5 in \cite{Fu}),
the differential equation holds even if the initial
condition $f$ is only in $L^1$.
In this case the convergence $f_t\to f$, as $t\to0+0$, holds only in $L^1$.
We remark that $T_t$ is also a contraction on $L^\infty$, by duality.

Now we restrict the dynamics to $\Sigma=\Sigma_N$.
Repeating the general construction with  $\bR^N$ replaced by $\Sigma_N$,
we obtain
the corresponding generator $L^{(\Sigma)}$ and the semigroup $T^{(\Sigma)}_t$.

To establish the relation between $L$ and $L^{(\Sigma)}$, we first
define  the symmetrized version of $\Sigma$
$$
\wt \Sigma:=
\bR^N\setminus \Big\{ \bx\; : \; \exists i\ne j \quad \mbox{with} \quad x_i=x_j \Big\}.
$$
Denote $X:= C_0^\infty(\wt \Sigma)$.
The key information is
that $X$ is dense in $H^1(\bR^N, \rd\mu)$ which is equivalent
to the density of $X$ in $C_0^\infty(\bR^N,\rd\mu)$. We will check this property
below.  Then the general argument above directly applies if $\bR^N$
is replaced by $\wt\Sigma_N$ and it shows  that the generator $L$ is the
the same (with the same domain) if we start from $X$ instead of $C_0^\infty(\bR^N,\rd\mu)$
as a  core.

Note that both $L$ with $L^{(\Sigma)}$ are local operators
and $L$ is symmetric with respect to the permutation of the variables.
For any function $f$ defined on $\Sigma$, we define its symmetric
extension onto $\wt \Sigma$ by $\wt f$. Clearly $L\wt f = \wt{ L^{(\Sigma)} f}$
for any $f\in C_0^\infty (\Sigma)$.
Since the generator  is uniquely
determined by its action on its core, and the generator uniquely determines
the dynamics, we see that
for any $f\in L^1(\Sigma, \rd\mu)$, one can determine $T^{(\Sigma)}_t f$
by computing $T_t \wt f$ and restricting it to $\Sigma$. In other
words, the dynamics \eqref{dy} is well defined when restricted to $\Sigma=\Sigma_N$.

Finally, we have to prove the density of $X$ in $C_0^\infty(\bR^N,\rd\mu)$,
i.e., to show that if $f\in C_0^\infty(\bR^N)$, then
there exists a sequence $f_n\in C_0^\infty(\wt\Sigma)$
such that $\cE(f-f_n, f-f_n)\to 0$. The structure of $\wt\Sigma$ is complicated
since in addition to the one codimensional coalescence hyperplanes $x_i=x_j$
(and $x_i=0$ in  case of $\Sigma^+$),
it contains higher order coalescence subspaces with higher codimensions. We will
show the approximation argument in a neighborhood of a point $\bx$
such that $x_i=x_j$ but $x_i\ne x_k$ for any other $k\ne i,j$.
The proof uses the fact that the measure $\rd\mu$ vanishes
at least to first order, i.e., at least as $|x_i-x_j|$, around $\bx$, thanks to $\beta\ge 1$.
This is the critical case; the argument near higher order coalescence
points is even easier, since they have lower codimension and the
measure $\mu$ vanishes at even higher order.

In a neighborhood of $\bx$ we can change to local coordinates such that $r:= x_i-x_j$
remains the only relevant coordinate. Thus the task is equivalent to show
that any $g\in C_0^\infty(\bR)$ can be approximated by a sequence
$g_\e \in C_0^\infty(\bR\setminus \{0\})$
in the sense that
\be
 \int_\bR |g'(r)-g_\e'(r)|^2 |r| \rd r \to 0
\label{gappr}
\ee
as $\e\to0$. It is sufficient to consider only the positive semi-axis, i.e., $r>0$.
Extending the functions to  two dimensional radial functions,
$G(x) : = g(|x|)$, $G_\e(x) = g_\e(|x|)$, this statement is equivalent to
the fact  that a point in two dimensions has zero capacity.

\section{Bakry-Emery argument on a subdomain}\label{sec:BE}

The estimate \eqref{0.1} in Theorem \ref{thm2} is based on
the Bakry-Emery argument \cite{BE} for the dissipation of the Dirichlet form.
This method uses a lower bound on the Hessian of $\wt\cH$ and an integration by parts.
Since the dynamics is restricted to $\Sigma=\Sigma_N$, we need to check that
the boundary term in the integration by parts vanishes.

In our application, this argument will be used for the Hamiltonian
$\wt \cH$ (see \eqref{def:wth}) and its
generator $\wt L =\frac{1}{2N}\Delta -\frac{1}{2}(\nabla\wt\cH)\nabla$, but for
simplicity, we omit the tilde from the notation below.
With $h=h_t=\sqrt{q_t}$
a standard calculation (see (5.8) of \cite{ERSY} with somewhat different
notations) shows that
\be
\begin{split}
\pt_t \frac{1}{2N}\int_{\Sigma} (\nabla h)^2 e^{-N\cH} \rd \bx
& =\frac{1}{N} \int_\Sigma \nabla h \nabla \Big(  L h+ \frac{1}{2N}
h^{-1} (\nabla h)^2\Big)e^{-N\cH} \rd\bx \cr
& = \frac{1}{N}
\int_\Sigma \Big[ \nabla h  L \nabla h -\frac{1}{2} \nabla h (\nabla^2
\wt\cH)\nabla h
+   \frac{1}{2N} (\nabla h)\nabla[h^{-1} (\nabla h)^2]\Big]e^{-N\cH}
\rd\bx
\cr
& = \frac{1}{2N}\int_\Sigma\Big[
-\nabla h (\nabla^2 \cH)\nabla h - \sum_{i,j}\big(\partial_{ij}^2 h -
\frac{\pt_i h
\pt_j h}{h}\Big)^2\Big] e^{-N\cH} \rd\bx
\cr
&\le - \frac{1}{2N}\int_\Sigma
\nabla h (\nabla^2 \cH)\nabla h \, e^{-N\cH} \rd\bx
\end{split}\label{be}
\ee
assuming that the quantities in each step are well defined and
that the boundary term
\be
 \int_{\pt \Sigma} \pt_i h\; \pt_{ij}^2 h \; e^{-N\cH}=0
\label{bdry}
\ee
in the integration by parts in the third line  vanishes.
In \cite{ERSY} we argued with a somewhat specific form of $q$,
an information  not directly available here.

The rigorous proof in the general case uses a regularization and
a cutoff argument.
First we
regularize the function $q=q_t\in D(L)$, $t>0$,  by defining
$$
  q^{\e} (\bx) := \frac{ q(\bx) + \e}{1+\e}, \qquad h^\e : = \sqrt{q^\e},
$$
for some $\e>0$. This has the advantage that the derivatives
of $h^\e$ can be bounded by those of $q^\e$.
We consider a
cutoff function $\theta\in C_0^\infty(\Sigma)$ to be
specified later  and we insert $\theta$ in the calculation \eqref{be}.
Since $L$ is an elliptic operator
with smooth coefficients away from the boundary $\pt\Sigma$,
by standard parabolic regularity we know that $q$ and thus
$h$ are smooth functions inside $\Sigma$. Thus
each step in the cutoff version of \eqref{be} is
justified with an additional term coming from
the derivative hitting $\theta$ in the integration  by
parts. After repeating the steps in \eqref{be}, we obtain
\be
\begin{split}\label{bet}
\pt_t \frac{1}{2N}\int_{\Sigma}& \theta(\nabla h^\e)^2 e^{-N\cH} \rd \bx
 =\frac{1}{N} \int_\Sigma\theta \nabla h^\e \nabla \Big(  L h^\e+ \frac{1}{2N h^\e}
(\nabla h^\e)^2\Big)e^{-N\cH} \rd\bx \cr
&\le - \frac{1}{2N}\int_\Sigma
\theta\nabla h^\e (\nabla^2 \cH)\nabla h^\e \, e^{-N\cH} \rd\bx
- \frac{1}2{N}\int_\Sigma
\sum_{i,j} (\pt_j\theta)(\pt_i h^\e) (\pt_i\pt_j h^\e) \, e^{-N\cH} \rd\bx .
\end{split}
\ee
We now show that, by an appropriate choice of a sequence of cutoff functions,
the second term in \eqref{bet} vanishes.
We first define the set of higher order coalescences where
at least three point coincide as
$$
  Q : = \{ \bx \in \pt\Sigma\; : \; \exists \; i \;\;\mbox{s.t.}\;\; x_i =x_{i+1} =x_{i+2}\}.
$$
We remark that in case of Assumption I'  we formally introduce
$x_0=0$ to this definition, so that $Q$ will
include also three point singularities of the type $x_1=x_2=0$.
For any $\delta>0$ we define the set
$$
 Q_\delta: = \{ \bx \in  \Sigma\; : \; \mbox{dist}(\bx, Q)\le \delta\}
$$
is the $\delta$-neighborhood of the three-point singularity set within $\Sigma$.
Introduce an additional small positive parameter $\eta\ll \delta$.
We now choose the cutoff function $\theta$ of the form
$\theta=\theta_1\theta_2$, depending on the parameters $\delta$ and $\eta$, such that

\medskip

(i) $\theta_1(\bx)\equiv 1$ if $\mbox{dist}(\bx, \pt\Sigma)\ge 2\eta$,
$\theta_1(\bx)\equiv 0$ if $\mbox{dist}(\bx, \pt\Sigma)\le \eta$
and $|\nabla\theta_1|\le O(\eta^{-1})$;

\medskip

(ii) $\theta_2(\bx)\equiv 1$ if $\mbox{dist}(\bx, Q)\ge 2\delta$,
$\theta_2(\bx)\equiv 0$ if $\mbox{dist}(\bx, Q)\le \delta$
and $|\nabla\theta_2|\le O(\delta^{-1})$.

\medskip
\noindent
Here and  in the sequel we make the convention that
a quantity of order $\delta^k$  with some $k\in \bR$ (sometimes denoted by $O(\delta^k)$)
denotes a number that is comparable with
$\delta^{k}$ with implicit constants
that may  depend on $N$. However, $N$ is fixed in
this argument, so this dependence is irrelevant.
Similar convention holds for $O(\eta^k)$.

We state two estimates on  the solution $q_t$ of \eqref{dytilde} that will
be proven at the end of the section.

\begin{lemma}\label{parreg} Assume that $q_0\in L^\infty$.
Then the solution $q_t$ of \eqref{dytilde} satisfies
a uniform supremum bound on the closure of $\Sigma$,
\be
\sup_{t\ge 0}\sup_{\bx \in \overline{\Sigma}} q_t(\bx) < \infty.
\label{linfty}
\ee
Furthermore, $q_t$ is regular away from the higher order coalescence singularities
with the estimate
\be
\sup \Big\{    |\nabla^k q_t(\bx)| \; : \; \bx \in \Sigma\cap K,
\; \mbox{dist}(\bx, Q)\ge \delta
\Big\} \le C(t, k, N, K) \delta^{-k}
\label{derest}
\ee
where $K$ is a compact set and the
constant depends only on the indicated parameters.
In particular, $q_t$ is regular up to the boundary $\pt\Sigma\setminus Q_\delta$, i.e.,
at the two-point coalescence points
away from higher order coalescences.
\end{lemma}

Using this lemma, we can treat the second term on the r.h.s. of \eqref{bet}.
We split the integration into two regimes. 
First we consider the regime where $\theta_2 \nabla \theta_1 \ne0$, i.e.,
an $(2\eta)$-neighborhood of  $\pt\Sigma\setminus Q_\delta$.
On this set we note 
the local density  scales at as $\eta^\beta$,
thanks to the term $|x_i-x_j|^\beta$ in $e^{-N\cH}$. Thus the
measure of the support of $\nabla \theta$ near  $\pt\Sigma\setminus Q_\delta$ 
scales as $\eta^{1+\beta}$,
while $|\nabla \theta|\le C \eta^{-1}$ (assuming $\eta\leq \delta$).
 Since \eqref{derest}
guarantees that the derivatives of $h^\e$
remain locally bounded (with a bound depending on $\e, \delta, t$ and $N$), the
boundary term near  $\pt\Sigma\setminus Q_\delta$ vanishes as $\eta\to 0$.

To estimate the integral on the support of $\nabla \theta_2$, i.e.,
on a subset of $Q_{2\delta}$,
we use that $\theta$ can be replaced with $\theta_2$ after taking the $\eta\to0$ limit.
Since we have
$|\nabla\theta_2| = O(\delta^{-1})$ and
$|\nabla^k h_t^\e|\le
C_\e|\nabla^k q_t^\e|\le C_{\e, t, N}\delta^{-k}$
with $k=1,2$, the integrand scales
at most $\delta^{-4}$. Since the
local density  scales at least as $\delta^{3\beta}$ due to
a factor of the type
$|x_i-x_{i+1}|^\beta|x_{i+1}-x_{i+2}|^\beta|x_i-x_{i+2}|^\beta$,
the total measure of $Q_{2\delta}$ is of order $\delta^{2+3\beta}$.
Hence the integral on $Q_{2\delta}$ scales at most as $\delta^{2+3\beta-4}\le \delta$
in the $\delta$ parameter and therefore
the contribution of the neighborhood
of higher order singularities to the second term
in \eqref{bet} vanishes as $\delta\to0$.

After having removed $\theta$ and the second term from  \eqref{bet},
we let $\e\to0$   and
this gives the desired result \eqref{0.1}.

To complete the argument,
finally, we need to prove Lemma \ref{parreg}.

\bigskip

{\it Proof of Lemma \ref{parreg}.}  The bound \eqref{linfty} follows immediately, since
$q_0\in L^\infty$ and the semigroup $T_t$ a contraction in $L^\infty$
(see Appendix \ref{sec:defdyn}).

\medskip

The second statement of Lemma  \ref{parreg}  follows
from a standard regularization  argument for a typical two-point singularity at $x_i=x_{i+1}$
that was already outlined in \cite{ESY4}.
Fix a point $\bx^* \in \pt\Sigma\setminus Q_\delta$
and assume that $x_i^* = x_{i+1}^*$, but  for all
other pairs $|x_j^*-x_{j+1}^*|\ge \delta$. We remark that the neighborhood of two (or more)
independent singularities, e.g., $x_i=x_{i+1}$ and $x_j = x_{j+1}$,
$|i-j|\ge 2$, can be treated similarly by
applying the same regularization argument separately. We omit
these details here.

Let $B$ be a neighborhood  of size  $O(\delta)$ around $\bx^*$.
Choose a local coordinate system $\Phi(\bx)=(u,\by)\in \bR_+\times
\bR^{N-1}$ in $B$ such that $u=\frac{1}{2}(x_{i+1}-x_i)>0$.
Within $\Phi(B)$, we can write
$$
\wt L = \frac{1}{4N} \Big[ \pt_u^2 + \frac{\beta}{u}\pt_u\Big] + L_{reg},
$$
where $L_{reg}$ is an elliptic operator with second derivatives
in the $\by$ variables and with coefficients  regular on the
scale $\delta$ (since all other singularities are at least at a distance
$O(\delta)$ away from $\Phi(B)$).

For the $\beta=1$ case,
by introducing a function $\wh q_t (a,b,\by):= q_t (\sqrt{a^2+b^2}, \by)$
of $N+1$ variables, we see that $\wh q_t$ satisfies $\pt_t \wh q_t = \wh L \wh q_t$,
where
$$
\wh L = \frac{1}{N} \big[ \pt_a^2 + \pt_b^2\big] + L_{reg},
$$
i.e., $L$ becomes an elliptic operator $\wh L$ with bounded and regular coefficients
in the new variables. A similar
transformation is possible for any integer $\beta \ge 1$, where $u$
is considered as the radial part of a $(\beta+1)$-dimensional variable.

We claim  that the singular point $u=0$  becomes a removable
singularity in the variables $(a,b)$ around $(0,0)$.
Note that the singular set is a  codimension two
subspace in the $(a,b,\by, t)$ space-time coordinate system
which becomes a line segment in the $(a,b,t)$ space-time system if
we disregard the variable $\by$. Note  that $\by$ plays no role in this argument since
every coefficient is regular in $\by$.
The parabolic equation  $\pt_t \wh q_t = \wh L \wh q_t$ holds
in a strong sense away from the origin $(a,b)=(0,0)$ in these
two variables, and moreover $\wh q_t$ is bounded
by \eqref{linfty}.
We can thus apply  Theorem II of \cite{Ar} with $p=2$, $r=\infty$
to see that $\wh q_t$ must coincide with the regular solution obtained
by using the fundamental solution to the equation in a small
space-time neighborhood of the singular set. This proves
that $\wh q_t$, and hence $q_t$,
is a smooth function up to the boundary $\pt\Sigma\setminus Q_\delta$.

To obtain the quantitative estimate \eqref{derest}, we consider the regularity of
the coefficients of $L_{reg}$. Due to the special
structure of $\cH$, every term in $L = \frac{1}{2N}\Delta-\frac{1}{2}(\nabla\cH)\nabla$
is either regular on any small scales, or
it scales as $\mbox{(length)}^{-2}$. Since the neighborhood $B$ is
at least at distance $O(\delta)$ away from the other singularities,
the coefficients of $L_{reg}$ are regular on scale $\delta$.
Therefore the solution $q_t$ is regular
on scale $\delta$ on $B$ and this gives the $\delta$-scaling of
the estimate \eqref{derest}. This completes the proof of Lemma
\ref{parreg}. \qed

\bigskip

{\it Acknowledgement.} The authors thank Alice Guionnet for 
pointing out some errors  in the preliminary version of
the manuscript.

\end{document}